\documentstyle[sprocl,epsfig]{article}

\def\be{\begin{eqnarray}}
\def\ee{\end{eqnarray}}
\def\bea{\begin{array}}
\def\eea{\end{array}}
\def\bei{\begin{itemize}}
\def\eei{\end{itemize}}

\begin{document}

\title{THE INFRARED LIMIT OF THE QCD DIRAC SPECTRUM AND APPLICATIONS OF
CHIRAL RANDOM MATRIX THEORY TO QCD}

\author{J.J.M.  Verbaarschot}

\address{Department of Physics and Astronomy, SUNY Stony Brook, \\
Stony Brook, NY 11790, USA,\\
e-mail: verbaarschot@nuclear.physics.sunysb.edu}

\maketitle\abstracts{
In the first part of these lectures we discuss the infrared limit of the 
spectrum of the QCD Dirac operator. We discuss the global symmetries of the 
QCD partition function and show that the Dirac spectrum near zero virtuality is 
determined by the pattern of spontaneous chiral symmetry breaking of a QCD-like
partition function with additional bosonic valence quarks and their 
super-symmetric partners. We show the existence of an energy scale below which 
the fluctuations of the QCD Dirac spectrum are given by a chiral Random Matrix 
Theory (chRMT) with the global symmetries of the QCD partition function. 
Physically, for valence quark masses below this scale the partition function is 
dominated by the zero momentum modes. In the theory of disordered systems, this
energy scale is known as the Thouless energy. In  the second part of these 
lectures we discuss chRMT as a schematic  model for the QCD partition function
at nonzero temperature and chemical potential.  We discuss novel features 
resulting from the non-Hermiticity of the Dirac operator. The analysis by 
Stephanov of the failure of the quenched approximation, the properties of 
Yang-Lee zeros, as well as the phase diagram of the chRMT partition function 
are discussed. We argue that a localization transition does not occur
in the presence of light quarks. Several results will be derived in full 
detail. We mention the flavor symmetries of the QCD Dirac operator for two 
colors, the calculation of the valence quark mass dependence of the chiral 
condensate  and the reduction of the chRMT partition function to
the finite volume partition function.}

\section{Introduction}
The two main phenomena that characterize the low-energy limit of 
QCD are confinement and chiral symmetry breaking. Because of 
confinement and the spontaneous breaking of chiral symmetry 
the low-lying states of QCD are given by the corresponding Goldstone bosons.
At low temperatures we expect that the thermodynamics of
the QCD partition function is that of a gas of (almost) massless
pions. Mainly because of lattice simulations
\cite{DeTar,Ukawa,Smilref,Karsch} we know that chiral symmetry is restored
above a critical temperature of about 140 $MeV$. The order parameter
of the chiral phase transition is the chiral condensate which through
the Banks-Casher formula \cite{BC}
is directly related to the spectral density of the
QCD Dirac operator. This is our main motivation for detailed
studies of QCD Dirac spectra.

One of the questions we wish to address in these lectures is to what extent
the infrared limit of the Dirac spectrum can be obtained analytically. 
After all, the infrared limit of the QCD partition function is determined by 
 the chiral Lagrangian describing Goldstone modes
that become noninteracting at zero momenta. 
On the other hand we know from many studies of complex systems that 
the correlations of eigenvalues on the scale of the individual level
spacings are universal and are given by Random Matrix Theory.
More precisely, this has been formulated as the Bohigas-Giannoni-Schmit
conjecture \cite{bohigas}
which states that spectral correlations of a classically
chaotic system are given by Random Matrix Theory 
(see Guhr et al. \cite{HDgang} for a comprehensive review).

The natural question one may ask is whether one can reconcile the
universal nature of the chiral Lagrangian with the universal spectral
correlations given by RMT. The first hint for the existence of 
a close connection came from the work of Leutwyler and Smilga \cite{LS}
who found that the existence of a low-energy chiral Lagrangian imposes
an infinite family of constraints on the spectrum of the Dirac operator.
On the other hand, in the field of Random Matrix Theory it has been known
for a long time that the spectral correlations can be described by
a Goldstone manifold arising from the spontaneous breaking of symmetry
between the advanced and retarded Greens functions 
\cite{Wegner,Stone,Efetov,VWZ}.  
The real breakthrough in this problem
came with the introduction of valence quarks
 \cite{Minn94,Osbornprl,Osborn,OTV,DOTV}. The idea is
\cite{Morel} to study a partition function 
that, in addition to the usual sea-quarks, 
contains a valence quark and its bosonic super-partner. 
After differentiation 
with respect to the source term and equating the fermionic and bosonic
valence quark masses, the determinants with the valence
quark masses cancel and we have access to the Dirac spectrum weighted
by the standard QCD action.
As is the case for the usual chiral Lagrangian, at low energies the
partition function involving both sea quarks and valence quarks is 
completely determined by the scheme of
chiral symmetry breaking and Lorentz-invariance and has been analyzed in
the context of partially quenched chiral perturbation theory 
\cite{pqChPT,Sharpe}.
However, instead
of constraints on QCD-Dirac spectra such as were obtained in the
work by Leutwyler and Smilga \cite{LS},
we are now in a position obtain an exact analytical expression
for the Dirac spectrum in the domain of applicability of this so called
partially quenched chiral perturbation theory. 

As is the case for the usual chiral perturbation theory \cite{GL}, one can
distinguish an important mass-scale. For sufficiently large volumes
such that the contributions of the non-Goldstone modes are suppressed 
and valence quark masses 
\be
   m_v \ll \frac {F^2}{\Sigma L^2} 
\label{thouless}
\ee
(the pion decay constant is denoted by $F$, $\Sigma$ is the chiral 
condensate and $L$ is the length of the box)
the contribution of the nonzero momentum modes factorizes from the 
partition function and
its mass dependence is given by the
contribution from the zero momentum modes alone 
\cite{vPLB,Trento,Osbornprl,Osborn,OTV,DOTV}.
However, QCD is not the only theory with this zero momentum sector. Any theory
with the same pattern of spontaneous symmetry breaking 
as QCD can be reduced to this zero momentum
limit. In particular, chiral Random Matrix Theory \cite{SVR,V}, with randomly 
distributed matrix elements of the Dirac operator is such a theory. 
In this case
the strong interactions required to achieve a spontaneous breaking of 
chiral symmetry result from strongly non-local random interactions.

Such a picture is well-know from the theory of disordered 
mesoscopic systems (a number of recent reviews
are available \cite{HDgang,Beenreview,Montambaux}).
In this case, the scale below which Random Matrix Theory is valid is known
as the Thouless energy, which is the inverse diffusion time
of an electron through the sample \cite{Altshuler}
\be
E_c = \frac {\hbar D}{L^2},
\label{thoulessm}
\ee
where $D$ is the diffusion
constant and $L$ is the linear dimension of the box. 
The domain with $\delta E \ll E_c$ is known as the ergodic domain whereas
the domain $E_c \ll\delta E \ll \hbar/\tau_e $ (with $\tau_e$ is the elastic
scattering time) is called the diffusive domain or the Altshuler-Shklovskii
domain.
In the case of QCD, the diffusion process is that of a quark propagating
through the Yang-Mills fields in a 4+1 dimensional space time.
The interpretation of spontaneous chiral symmetry
breaking as a Mott transition was
given earlier in \cite{shuryak}. By analogy with the
Kubo formula, $\Sigma$ plays the role of the conductivity 
\cite{shuryak}.

By now, this  picture of the QCD gauge field configurations as a disordered
system has been investigated in many lattice QCD simulations
\cite{Halasz,vPLB,Trento,Tilo,Ma,Tilomore,many,Guhr-Wilke,Tilomass}
\cite{tilomark,tilochem,markum,Dam3d,DamSU3,TiloSU3,Berg,Janik,Hehl,Tilo99} 
and several
instanton liquid simulations \cite{SVR,Vinst,Osbornprl,Osborn}. 
In particular,
it has been confirmed that the scale below which chiral Random Matrix Theory is
valid is given by the Thouless energy (\ref{thouless})
\cite{Osbornprl,Osborn,many,TiloSU3,Tilo99}. 
It turns out that eigenvalue correlations
are given by chRMT all the way up to this scale.
Beyond this energy scale one has to take into account the contributions
of the nonzero momentum modes. This can be done simply to one-loop order in 
chiral perturbation theory. In that way, one obtains analytical results
for the valence quark mass dependence \cite{GolLeung,OTV}
of the chiral condensate in the Altshuler-Shklovskii domain. 

In the first part of these lectures we will  show that chRMT is an 
$exact$ theory for the spectrum of the QCD Dirac operator in the
ergodic domain. This is achieved by showing that in the range
(\ref{thouless}) the valence quark mass dependence of the chiral
condensate obtained from the low-energy limit of the QCD partition
function coincides 
with the valence quark mass dependence obtained
from chiral Random Matrix Theory.
In the second part of these lectures we take a different route and use
chRMT as a $schematic$ model of the 
chiral phase transition 
\cite{JV,Tilo1,Stephanov1,Stephanov,Nowak,Halaszyl}. 
In that
case we obtain only $qualitative$ results for the spectrum of the QCD Dirac
operator and the chiral phase transition. For example, such models
are the random matrix equivalent of  a 
Landau-Ginzburg functional and give rise to mean field critical exponents.
At the tricritical point where the upper critical dimension is three they
provide us with a reliable description of the the infrared degrees of
freedom \cite{phase}.
The advantage of using random matrix models as opposed to a Landau-Ginzburg
functional is that one has access to the Dirac spectrum. In particular,
at nonzero chemical potential, this has provided us with valuable insights
in the effects of non-Hermiticity
on the properties of the QCD
partition function \cite{Stephanov,Feinberg-Zee}. The common ingredient
of chRMT and QCD at nonzero chemical potential is 
that the phase of the fermion determinant
due to the non-Hermiticity makes Monte Carlo simulations impossible. Since
the chRMT partition function 
can be evaluated analytically, we are in a position to analyze some
of these problems in great detail. In this way, a complete 
analytical understanding 
of the failure of the quenched approximation at nonzero chemical potential has
been obtained \cite{Stephanov}.

In the first lecture (section 2) we summarize the main properties of
the Euclidean QCD partition function. We discuss in detail the 
global symmetries and give a brief review of lattice QCD (see lectures
by Tom Blum \cite{blum} 
for more details). In the second lecture we analyze the infrared 
limit of the QCD partition function (section 3). We discuss the domain of
validity of the zero momentum limit of  the effective chiral
partition function and introduce the partially quenched chiral
partition function. An important point of this lecture is the discussion of
the integration manifold. Explicit calculations of the valence
quark mass dependence of the chiral condensate that rely on 
the structure of this Riemannian super-manifold are presented in section 4.
Results are derived in detail both by perturbative and 
non-perturbative methods. 
In the third lecture we introduce chiral Random Matrix Theory. We discuss
its main properties and emphasize the importance of universality (see section
5). In the second half of this lecture we present lattice QCD results for 
the statistical properties of Dirac eigenvalues 
and compare the results with random matrix
theory. In the last lecture we introduce chiral Random Matrix Theory
as a schematic model for the chiral
phase transition. In the framework of this model, we discuss chiral
symmetry breaking, quenching at nonzero chemical potential, Yang-Lee zeros
in the complex chemical potential plane, the structure of Dirac spectra
at nonzero chemical potential and the phase diagram in the mass, chemical
potential and temperature plane.
Finally, we present general arguments that a localization transition does
not occur in QCD with light quarks. Concluding remarks are made in section 8.

\section{The QCD Partition Function}
The QCD partition function describing strong interactions in a
box of volume $V_3=L^3$ can be expressed in terms of the eigenvalues of the
QCD Hamiltonian $E_k$ as
\be
Z^{QCD} = \sum_k e^{-\beta E_k},
\ee
where $\beta$ is the inverse temperature. At low temperatures $(\beta 
\rightarrow \infty)$ the partition function is dominated by the lightest
states of the theory, namely the vacuum state, with an energy density of 
$E_0/V_3$, and massless excitations thereof. 
The partition function $Z_{QCD}$
can be rewritten as 
Euclidean functional integral over the nonabelian gauge fields $A_\mu$
\be
Z^{QCD} = \int d A_\mu \prod_{f=1}^{N_f} \det (D + m_f) e^{-S_{YM}},
\label{ZQCD}
\ee
where $S_{YM}$ is the Yang-Mills action.
We assume that this partition function is properly regularized, for  
example, by a lattice regularization.
The gauge fields $A_\mu$ are nonabelian gauge fields in Euclidean space
time of volume $V = L^3 \beta$. In terms of the generators,  $T_a$, 
of $SU(N_c)$ they can be written as
\be
A_\mu = A_\mu^a \frac {T_a}2,
\ee
The anti-Hermitean Dirac operator is defined by
\be
D = \gamma_\mu (\partial_\mu + i A_\mu),
\label{Dirac}
\ee
where the $\gamma_\mu$ are the Euclidean Dirac matrices with 
$\{\gamma_\mu, \gamma_\nu\} = 2\delta_{\mu\nu}$. In these lectures we will
use the chiral representation of the gamma matrices (with diagonal 
$\gamma_5$).

\subsection{Symmetries of the QCD Partition Function}

It is well-known that the QCD action is greatly 
constrained by gauge symmetry, Euclidean Poincar\'e invariance 
and renormalizability. These symmetries determine the structure
of the Dirac operator (\ref{Dirac}). In this section we will discuss
the global symmetries of the Dirac operator. They play an essential role
in its spectral properties in the deepest infrared. In particular,  
the chiral symmetry, the flavor symmetry and
the anti-unitary symmetry of the continuum Dirac operator are discussed.

The chiral symmetry, or the $U_A(1)$ symmetry, can be expressed in terms of
the anti-commutation relation
\be
\{\gamma_5, D\} =0.
\label{ua1}
\ee
This implies that all nonzero eigenvalues occur in pairs $\pm i\lambda_k$ with
eigenfunctions given  by $\phi_k$ and $\gamma_5 \phi_k$. If $\lambda_k = 0$ the
possibility exists that $\gamma_5 \phi_k \sim \phi_k$, so that $\lambda_k=0$ 
is an unpaired eigenvalue. According to the Atiyah-Singer theorem, 
the total number of such zero eigenvalues
is a topological invariant, i.e., it does not change under
continuous transformations of the gauge field configuration.
Indeed, this possibility is realized by  the field of an instanton 
which is a solution
of the classical equations of motion. On the other hand, it cannot
be excluded that $\lambda_k = 0$ while $\phi_k$ and $\gamma_5\phi_k$ are
linearly independent. However, this imposes additional constraints on
the gauge fields which will be violated by infinitesimal deformations.
Generically, such situation does not occur.

In a decomposition according to the total number of topological zero modes,
the QCD partition function can be written as
\be
Z^{QCD} = \sum_\nu e^{i\nu\theta} Z_\nu^{QCD},
\label{ZQCDdet}
\ee
where
\be
Z^{QCD}_\nu = \langle\prod_f m^{\nu}_f \prod_k 
(\lambda_k^2 + m^2_f)
\rangle_\nu.
\ee
Here, $\langle \cdots \rangle_\nu$ denotes the average over gauge-field
configurations with topological charge $\nu$ weighted by the Yang-Mills
action. In we introduce  complex conjugated right-handed and left-handed 
masses we observe that the $\theta$ dependence of the QCD
partition function is only through the combination $m e^{i\theta/N_f}$.
 
A second important global symmetry is the flavor symmetry. This symmetry
can be best explained by writing the fermion determinant in the QCD partition
function as a functional  integral over Grassmann variables,

\be
\prod_f \det(D+m_f) = \int d \psi d \bar \psi e^{\int d^4x
\sum_{f=1}^{N_f} \bar \psi^f (D+ m_f) \psi^f}.
\ee
In a chiral basis with $\psi_R = \gamma_5 \psi_R$
and $\psi_L = -\gamma_5 \psi_L$, this can be rewritten as
\be
\prod_f \det(D+m_f) = \int d \psi d \bar \psi e^{\int d^4 x[\sum_{f=1}^{N_f}
\bar \psi^f_R D \psi^f_L + \bar \psi^f_L D \psi^f_R +
\bar \psi^f_R  m_f \psi^f_R+ \bar \psi^f_L  m_f \psi^f_L]}.\nonumber \\
\ee
For $m_f = 0$ we have the symmetry
\be
  \psi_L \rightarrow U \psi_L, \quad
\bar \psi_R \rightarrow \bar\psi_R U^{-1}, \nonumber \\
\bar  \psi_L \rightarrow \bar \psi_LV^{-1}, \quad
\psi_R \rightarrow V \psi_R.
\ee
The only condition to be imposed on $U$ and $V$ is that their inverse exists.
If the number of lefthanded modes is equal to the number of right-handed modes
we thus have an invariance under $Gl(N_f)_R \times Gl(N_f)_L$. However, if
the number of left-handed modes is not equal to the number of right-handed
modes, the axial-symmetry group
is broken to an $Sl(N_f)$ subgroup whereas the vector
symmetry with $U=V$ remains unbroken.
For $m=0$ the flavor symmetry is thus broken explicitly to 
$Gl_V(N_f)\times Sl_A(N_f)$ by instantons or the anomaly.

What is much more important, though, is the spontaneous breaking of
the axial flavor symmetry. From lattice QCD simulations and phenomenological
arguments we know that the expectation value
\be
\langle \bar \psi \psi \rangle =
\langle \bar \psi_R \psi_R \rangle +
\langle \bar \psi_L \psi_L \rangle\approx (240\, MeV)^3
\ee
in the vacuum state of QCD instead of the symmetric possibility
$\langle \bar \psi \psi \rangle = 0$.
Phenomenologically, this is known 
because of the presence of Goldstone modes. The pions are much lighter than
the $\sigma$ mesons. The spontaneous breaking of the axial symmetry  
also follows from the absence of
parity doublets. For example, the pion mass and the $\delta$ (or $a_0$) mass
are very different ($m_\pi = 135 MeV$ and $m_\delta = 980 MeV$).

On easily verifies  
that $\langle \bar \psi \psi \rangle$ is only invariant for $U=V$.
The vacuum state thus breaks the chiral symmetry down to $Gl_V(N_f)$.
Notice that only the axial symmetries are broken. This is in agreement
with the Vafa-Witten theorem \cite{Vafa}
which states that vector symmetries cannot
be broken in vector-like theories such as QCD. 
We also observe that the $complete$
axial group is broken. The reasons behind this maximum breaking 
\cite{Shifman-three} of chiral symmetry are less well understood.
The Goldstone manifold is given by the maximum Riemannian submanifold
of $Sl_A(N_f)$ which is the usual manifold $SU_A(N_f)$. The complex 
extension of $SU(N_f)$ does not give rise to additional conserved currents,
and therefore, the total number of Goldstone modes remains the same
(we thank J.C. Osborn for this remark).

For three or more colors the gauge fields are not related by complex
conjugation and there are no anti-unitary symmetries. However, the situation
is more interesting for $N_c=2$ which will be discussed in the next section.

   \subsection{Special Role of $N_c =2$}

The group $SU(2)$ is the only nontrivial special unitary group
that  is pseudoreal. As will be shown in this section, 
the  consequence is that for 
$N_c= 2$ the symmetry group of the QCD partition function in
the chiral limit is enlarged to $U(2N_f)$.

As before, the fermionic action is obtained by writing the determinant 
in the partition
function as a Gaussian integral over the Grassmann fields $\phi$ and $\bar 
\phi$. Using the chiral representation of the $\gamma$ matrices
with
\be
\gamma_\mu = \left ( \begin{array}
{cc} 0 & \hat \sigma_\mu  \\
                   \hat\sigma_\mu^+ & 0 \end{array} \right ),
\ee
and $\hat\sigma_\mu = (1, i\sigma_k)$ 
with $\sigma_k$ the Pauli $\sigma$-matrices,
the fermionic action can be written as
\be
S_F = \int d^4 x \sum_{f=1}^{N_f} 
\left( \bea{c} \bar \phi_R \\ \bar \phi_L \eea \right )
\left ( \bea{cc} m_f &\hat\sigma_\mu (\partial_\mu + i A_\mu) \\
        \hat\sigma_\mu^+ (\partial_\mu + i A_\mu) & m_f \eea \right )
\left( \bea{c}  \phi_R \\  \phi_L \eea \right ).\nonumber \\
 \ee
For $N_c = 2$, we have that $A_\mu^T = -\tau_2 A_\mu  \tau_2$ (with 
$\tau_2 = \sigma_2$ in color space). Combining this
with the relation $\hat\sigma_\mu^* = \sigma_2 \hat\sigma_\mu \sigma_2$ we find
\be
\bar\phi_L^f \hat\sigma_\mu^+ (\partial_\mu + i A_\mu) \phi_R^f =
 \phi_R^f \sigma_2\tau_2 \hat\sigma_\mu (\partial_\mu + i A_\mu) 
\sigma_2\tau_2 \bar \phi_L^f ,
\ee
where we have used that $\partial_\mu$ is anti-Hermitean and that the
fermion fields are anti-commuting Grassmann variables.
The fermionic action can thus be rewritten as
\be
S_F(m_f = 0) = \int d^4 x \sum_{f=1}^{N_f} 
\left( \bea{c} \bar \phi_R^f \\ \sigma_2 \tau_2 \phi_R^f \eea \right )
\left( \bea{cc}\hat\sigma_\mu (\partial_\mu + i A_\mu)& 0 \\
        0& \hat\sigma_\mu (\partial_\mu + i A_\mu) \eea \right )
\left( \bea{c}  \phi_L^f \\  \sigma_2\tau_2 \bar \phi_L^f \eea \right ).
\nonumber\\
 \ee
Obviously the symmetry group is $Gl(2N_f)$. If the number of left-handed
modes is not equal to the number of right-handed modes because of the
anomaly, an axial $U(1)$ is broken explicitly as for three or more
colors. 

The mass term is given by
\be
S_m &=& \int d^4 x \sum_{f=1}^{N_f} \left [
\left( \bea{c}  \phi_L^f\\  \sigma_2\tau_2 \bar \phi_L^f \eea \right )
\left ( \bea{cc} 0& -m_f\sigma_2\tau_2  \\
        m_f\sigma_2\tau_2 & 0 \eea \right ) 
\left( \bea{c}  \phi_L^f \\  \sigma_2\tau_2 \bar \phi_L^f \eea \right )
\right . \nonumber \\
&+& \left .
\left( \bea{c} \bar \phi_R^f \\ \sigma_2\tau_2  \phi_R^f \eea \right )
\left ( \bea{cc} 0& m_f\sigma_2\tau_2  \\
        -m_f\sigma_2\tau_2 & 0 \eea \right ) 
\left( \bea{c} \bar \phi_R^f \\ \sigma_2\tau_2  \phi_R^f \eea \right )
\right ] .
\ee
Also in this case we expect maximum spontaneous chiral 
symmetry breaking consistent
with the Vafa-Witten theorem. This means that only the subgroup of $U(2N_f)$
that leaves both $\bar \phi_R \phi_R$ and $\bar \phi_L \phi_L$ invariant 
remains unbroken. Therefore, only the subgroup  that leaves
\be
\left( \bea{c}  \phi_L^f\\ \bar \sigma_2\tau_2 \bar \phi_L^f \eea \right )
\left ( \bea{cc} 0& -\sigma_2\tau_2  \\
        \sigma_2\tau_2 & 0 \eea \right ) 
\left( \bea{c}  \phi_L^f \\  \sigma_2\tau_2 \bar \phi_L^f \eea \right )
\ee
invariant remains unbroken ($\bar \phi_R \phi_R$ is invariant under
the same transformations). This is the symplectic group $Sp(N_f)$.
The Goldstone manifold is thus given by the coset
$SU(2N_f)/Sp(N_f)$.

For $N_c=2$ the Dirac-operator has the anti-unitary symmetry
\be
[KC\tau_2, iD] = 0,
\label{anti1}
\ee
where $K$ is the complex conjugation operator, $C= \gamma_2\gamma_4$ is the 
charge conjugation matrix.
Because
\be
(KC\tau_2)^2= 1
\ee
we can always find a basis such that the Dirac matrix is real for any $D$.
The proof is along the same lines as the proof that time reversal symmetry
results in real matrix elements for the Hamiltonian in quantum mechanics.

The anti-unitary symmetry would be  violated in the presence of gauge fields
$B_\mu$ coupling to the axial current (which are not present in QCD). 
In that case the Dirac operator
is given  by 
\be
\gamma_\mu \partial_\mu + i\gamma_\mu A_\mu+\gamma_\mu \gamma_5 B_\mu,
\ee
which, because of the absence of the  factor $i$ in the $B_\mu$-term 
does not 
satisfy the commutation relation (\ref{anti1}). In other words, 
the invariance of the Dirac 
operator  under (\ref{anti1}) follows from 
the condition $\gamma_\mu D \gamma_\mu = 2 D$ (in four dimensions). For $N_c 
\ge 3$ this condition does not affect the anti-unitary symmetries, and the
the matrix elements of the Dirac operator are 
arbitrary complex numbers both with and without the $B_\nu$ 
term \cite{sternpc}.

\subsection{Dirac Operator in a Chiral Basis}
It is instructive to write the Dirac operator in a chiral basis with
$n$ right-handed modes $\phi_k^R$ and $n+\nu$ left-handed modes $\phi_k^L$.
Using the anti-commutation relation (\ref{ua1}) one can easily derive that
\be
\langle \phi_k^R| D | \phi_l^R\rangle =
\langle \phi_k^R| \gamma_5 D \gamma_5 | \phi_l^R\rangle=
-\langle \phi_k^R| D | \phi_l^R\rangle
\ee
from which it follows that $\langle \phi_k^R| D | \phi_l^R\rangle = 0$. 
Similarly, $\langle \phi_k^L| D | \phi_l^L\rangle = 0$, resulting in a
Dirac Matrix with  the  structure
\be
D = \left ( \begin{array}{cc} 0 & T \\ -T^\dagger & 0 \end{array} \right ),
\label{Dmatrix}
\ee
where $T$ is an $n\times (n+\nu)$ matrix. From the inspection of the secular 
equation one concludes  that such matrix has always exactly $\nu$ zero 
eigenvalues. For example, the eigenvalues of the matrix 
\be
D = \left ( \begin{array}{ccc} 0 & a&b\\ -a^* &0&0\\-b^*&0&0
\end{array} \right ),
\ee
are equal to 0 and $\pm i \sqrt{a^*a+b^*b}$. 
The matrix structure (\ref{Dmatrix}) is at the basis of the Random Matrix Theory
to be discussed in section 4. A particular realization of a chiral basis is
a basis of instanton zero modes which has been used extensively in the 
instanton liquid model of the QCD vacuum \cite{instantons,shuryak}

  \subsection{Lattice QCD}
In lattice QCD, the 
gauge fields represented by  unitary matrices on the links of the lattice, i.e.
\be
U_{x,x+\mu} = e^{i \frac{T_a}2 A^a_\mu(x)},
\ee
where $\mu$ denotes a unit vector in the direction of $x_\mu$.
The gluonic lattice action of QCD is the plaquette action given by
\be 
S = \frac 12\sum_P (S_P + S_P^\dagger).
\ee
For example, the action of an elementary plaquette $1234$ is given  by 
\be
S_P = {\rm Tr} U_{12} U_{23} U^{-1}_{43}
U^{-1}_{14}.
\ee
This action is invariant under the local gauge transformations
\be
U_{kl} \rightarrow V_k^{-1} U_{kl} V_l.
\ee
The  fermion fields are represented by Grassmann valued fields on the sites.
Under gauge transformations they transform according to
\be
\psi_k \rightarrow U_{kl} \psi_l.
\ee

However, a lattice discretization of the Dirac action is 
more problematic. In a
naive lattice discretization we are faced with the fermion doubling
problem. This is highlighted by  
the Nielsen-Ninomiya theorem \cite{holger}, which states
that a lattice discretization of chiral fermions with local interactions
is not possible. 

There are two widely used discretizations of the fermionic action which
deal with the doubling problem at the expense of some of the symmetries
of the continuum QCD action: the 
Kogut-Susskind or staggered action and the Wilson action. The Kogut-Susskind
action is defined by

\be
S^{KS} &=& \frac 12 \sum_{n\,\mu} \eta_\mu(n) \bar\chi_n ( 
U_{n,n+\mu}\chi_{n+\mu}-  U_{n-\mu,n}^\dagger \chi_{n-\mu}) 
+m\sum_n \bar \chi_n \chi_n
\nonumber \\
&=& \bar \chi_m D^{KS}_{mn} \chi_n, 
\ee
where $\eta_\mu(n)$ is a phase factor resulting from the diagonalization of
the $\gamma$-matrices. For $m= 0$ and $n_1+n_2+n_3+n_4$ even, 
this action is invariant under 
the transformations
\be
\bar \chi_n \rightarrow e^{i\theta}   \bar \chi_n, \quad \quad
 \chi_{n+\mu} \rightarrow e^{-i\theta}    \chi_{n+\mu}.
\ee
For all $n_i$ we have the invariance
\be
\bar \chi_n \rightarrow e^{i\alpha}   \bar \chi_n, \quad  \quad
 \chi_n \rightarrow e^{-i\alpha}   \bar \chi_n.
\ee
For $m \ne 0$ this $U(1)\times U(1)$ symmetry is broken to the
second $U(1)$ symmetry only.  
If we organize the Kogut-Susskind Dirac matrix according to blocks with
even and odd   $n_1+n_2+n_3+n_4$, one obtains the same block structure as
in the chiral representation of the continuum Dirac operator
(\ref{Dmatrix}). This shows that
that the eigenvalues of $D^{KS}$ occur in pairs $\pm \lambda$. 

The staggered Dirac operator is anti-Hermitean with eigenvalues $i\lambda_k$
on the imaginary axis. The Dirac operator satisfies the sum-rule
\be
{\rm Tr} D^{KS} D^{KS} =\left . {\rm Tr} D^{KS} D^{KS} \right |_{U= 1} .
\label{sumks}
\ee
This follows from the observation that $U$ only occurs in the
combination $U U^{-1}$ in the left hand side of this equation. 
An important application of this sum rule is as a test
of 
 the accuracy of numerically calculated eigenvalues of
the Dirac operator.

The Wilson Dirac action is defined by
\be
S^W &=& -\frac 12 \sum_{n\, \mu} \bar \psi^f_n((r-\gamma_\mu) 
U_{n,n+\mu}\psi_{n+\mu}
 +(r+\gamma_\mu) U_{n-\mu,n} ^\dagger) \psi^f_{n-\mu}
+ \sum_n\bar \psi^f_n (m+4r) \psi^f_n \nonumber \\
&=& \bar \psi^f_n D^W_{nm} \psi^f_m.
\ee
The term proportional to $r$ was introduced by Wilson to remove the fermion
doublers. However, it also destroys the $U_A(1)$ symmetry of $D^W$, and the
eigenvalues do not occur in pairs $\pm \lambda$. On the other hand, this
action is invariant under the flavor group $SU_V(N_f)$.
The Wilson Dirac operator satisfies the Hermiticity relation
\be
D^{W \,\dagger} = \gamma_5 D^W \gamma_5
\ee
from which it follows that $\gamma_5 D^W$ is Hermitean. In the literature
Dirac spectra of both $D^W$ with eigenvalues scattered in the complex
plane \cite{Bardeen,Sailer,GHL}
and $\gamma_5 D^W$ with eigenvalues on the real axis 
\cite{gathip,Hip} have 
been studied (several reviews appeared recently \cite{negele,narayanan}).
Below we will restrict ourselves to the Hermitean Wilson
Dirac operator.  The Wilson Dirac operator satisfies the sum rule
\be
{\rm Tr} D^{W \,\dagger} D^{W} =\left . {\rm Tr} D^{W \, \dagger} D^{W} 
\right |_{U= 1} .
\label{sumwil}
\ee
Again this can be seen from the observation that $U$ only occurs in the
combination $U U^{-1}$ in the l.h.s. of this equation.

The anti-unitary symmetries of the Wilson Dirac operator are the same
as of the continuum Dirac operator. For $N_c =2 $ we thus have
\be
[CK \tau_2, iD^W] =0.
\ee
For staggered fermions only the eigenvalues of the $\gamma$ matrices 
appear in the Dirac operator. Since they are real they do not affect
the anti-unitary symmetry. For $N_c = 2$ we thus have

\be
[K\tau_2, D^{KS}] =0.
\label{ks-com}
\ee
However, $(K\tau_2)^2 = -1$, and it can be shown that it is always possible
to find a basis such that the matrix elements of $D^{KS}$ 
can be organized into real  quaternions for any gauge field
configuration. The 
eigenvalues of such matrix are scalar quaternions and are thus doubly 
degenerate. This can also easily be shown from (\ref{ks-com}). Because of
this relation, if $\phi$ is an eigenfunction with eigenvalue $\lambda$ then
$K\tau_2 \phi$ is also an eigenfunction with eigenvalue $\lambda$. But
\be
(\phi, K\tau_2 \phi)=(K\tau_2\phi, \phi)^*= ((K\tau_2)^2\phi, K\tau_2 \phi)
=-(\phi, K\tau_2 \phi),
\ee 
from which it follows that $\phi$ and $K\tau_2 \phi$ are linearly independent.
For a more detailed discussion of  the symmetries of the staggered
Dirac operator we refer to \cite{Stone,Hands,Lagae}.

    \subsection{The Chiral Phase Transition in QCD}
It is  expected that chiral symmetry will be restored above some
critical temperature. This is supported both by naive arguments
based on counting the total number of degrees of freedom and 
by lattice QCD simulations \cite{DeTar}. In the chiral limit,
the order parameter for the chiral phase transition is the
chiral condensate, $\langle \bar \psi \psi \rangle$.  For QCD with
three color and two massless quarks, it vanishes
above a critical temperature of about 140 $MeV$.
The nature of the phase transition is
still under dispute. It is not yet clear whether the transition is
a second order or a weak first order one \cite{DeTar}. 
The role of instanton field
configurations has to be clarified as well \cite{SS,negele}. 
However, with current progress
in computational resources, these questions should be answered in 
the near future. The only remaining fundamental problem is the
study of QCD at finite baryon density. Because of the phase of the
fermion determinant, this problem cannot be resolved by means
of Monte-Carlo simulations. Progress in this area requires new paradigms
which make it a particularly challenging area of active research.

\subsection{The Banks-Casher Relation and Microscopic Spectral Density}

The order parameter of the chiral phase transition,
$\langle \bar \psi \psi \rangle$,
is nonzero only below the critical temperature or a critical chemical
potential.
As was shown by Banks and Casher {\cite{BC}},
$\langle \bar \psi \psi \rangle$ is directly related to the eigenvalue density
of the QCD Dirac operator per unit four-volume
\be
\Sigma \equiv
|\langle \bar \psi \psi \rangle| =\lim\frac {\pi \langle {\rho(0)}\rangle}V,
\label{bankscasher}
\ee
where the spectral density of the Dirac operator with eigenvalues
$\{\lambda_k\}$ is defined by
\be
\rho(\lambda) = \langle \sum_k \delta(\lambda-\lambda_k) \rangle.
\ee
It is elementary to derive this relation.
The chiral condensate is defined as the logarithmic derivative
of the
the partition function (\ref{ZQCDdet}). For equal quark masses we obtain,
\be
\langle \bar \psi \psi \rangle  = -\lim\frac 1{VN_f}\partial_m \log Z^{QCD}(m)
=-\lim\frac 1V \langle \sum_k \frac {2m}{\lambda_k^2 + m^2}\rangle.
\label{preBC}
\ee
If we express the sum as an integral over the average spectral density,
and take the thermodynamic limit before the chiral limit, so that many
eigenvalues are less than $m$, we recover (\ref{bankscasher}). The order of the
limits in (\ref{bankscasher}) is important. First we take the
thermodynamic limit, next the chiral limit and, finally, the field theory
limit. As follows from (\ref{preBC}), the
sign of $\langle \bar \psi \psi \rangle $ changes if $m$ crosses the
real  axis.

An important consequence of the Bank-Casher formula (\ref{bankscasher})
is that the eigenvalues near zero virtuality are spaced as
\be
\Delta \lambda = 1/{\rho(0)} = {\pi}/{\Sigma V}.
\label{spacing}
\ee
This should be contrasted with the eigenvalue spectrum
of the non-interacting
Dirac operator. Then the eigenvalues are those of a free Dirac particle in a 
box with eigenvalue spacing equal to  $\Delta \lambda \sim 1/V^{1/4}$
for the eigenvalues near $\lambda = 0$.
Clearly, the presence of gauge fields leads to a strong modification of
the spectrum near zero virtuality. Strong interactions result in the
coupling of many degrees of freedom leading to extended states 
and correlated eigenvalues.

Because the eigenvalues near zero are spaced as $\sim 1/\Sigma V$
it is natural to introduce the microscopic variable
\be
u = \lambda V \Sigma,
\ee 
and the microscopic spectral density \cite{SVR}
\be
\rho_s(u) = \lim_{V\rightarrow \infty} \frac 1{V\Sigma} \langle
\rho(\frac u{V\Sigma})\rangle.
\label{rhosu}
\ee
We expect that this limit exists and converges to a universal function
which is determined by the global symmetries of the QCD Dirac operator.
The calculation of this universal function from QCD is the main 
objective of these lectures. 
We will calculate $\rho_s(u) $ both from the simplest theory in this
universality class which is chiral Random Matrix Theory (chRMT) and 
from the partial quenched chiral Lagrangian which describes the low-energy
limit of the QCD partition function. We find that the two results coincide
below the Thouless energy.

\subsection{Valence Quark Mass Dependence of the Chiral Condensate}

Instead of studying the Dirac spectrum it is often convenient
to consider the valence quark mass dependence of the chiral condensate.
In terms of the eigenvalues of the Dirac operator it is defined by
\cite{Christ,vPLB,Trento}
\be
\Sigma(m_{v};m_1,\cdots, m_{N_f})& =&
\frac 1V \sum_k \left \langle \frac 1{i \lambda_k + m_v}
\right \rangle \nonumber \\
&=&
\frac 1V
\int d\lambda \frac{\rho(\lambda; m_1,\cdots, m_{N_f})}
{i\lambda + m_v}.
\label{massdep}
\ee
Here,  $\langle \cdots \rangle$ denotes an average with respect to
the distribution of the eigenvalues.

The relation (\ref{massdep})
can then be inverted to give $\rho(\lambda;m_1,\ldots,m_{N_{f}})$.
As mentioned in \cite{OTV}, the spectral density
follows from the discontinuity across the imaginary axis,
\be
\left .{\rm Disc}\right |_{m_v = i\lambda}\Sigma(m_v)
= \lim_{\epsilon \rightarrow 0}
\Sigma(i\lambda+\epsilon) - \Sigma(i\lambda-\epsilon) = 2\pi \sum_k
\langle \delta(\lambda +\lambda_k)\rangle
= 2\pi \rho(\lambda),\nonumber\\
\label{spectdisc}
\ee
where we have suppressed the dependence on the sea-quark masses.

\section{Infrared Limit of the QCD Partition Function}
     \subsection{The Chiral Lagrangian}
For light quarks the low energy limit of QCD is well understood. It is
given by the chiral
Lagrangian describing the interactions of the pseudoscalar mesons. 
The reason is that pions are Goldstone bosons which
 are the only light degrees of freedom in a confining theory such as
QCD. To lowest order in the quark masses and the momenta, 
the chiral Lagrangian is completely dictated by chiral symmetry and 
Lorentz invariance. In the
case of $N_f$ light quarks with chiral symmetry breaking according to 
$SU_L(N_f) \times SU_R(N_f) \rightarrow SU_V(N_f)$ the so called
Weinberg Lagrangian is given by \cite{GaL}
\be
{\cal L}_{\rm
eff}(U)=\frac{F^2}{4} \; {\rm Tr} (\partial_\mu U \; \partial_\mu
U^\dagger)-\frac{\Sigma}{2} \; {\rm Tr} (\hat{\cal M} U+\hat{\cal
M} U^\dagger),
\ee
where $F$ is the pion decay constant, $\Sigma$ is the chiral condensate
and ${\cal M}$ is the quark mass matrix. 
The fields $U(x)$ are $SU(N_f)$ matrices 
parametrized as
\be
 U={\rm exp} (i\sqrt 2 \Pi_a t^a/F),
\ee
with the generators of $SU(N_f)$ normalized according to ${\rm Tr} t^a t^b
=\delta^{ab}$.
This chiral  Lagrangian has been used extensively
for  the description of pion-pion scattering amplitudes.

To lowest order in the pion fields the chiral Lagrangian
can be expanded as (for equal quark masses)
\be
{\cal L}_{\rm eff}(U) = \frac 12 \partial_\mu \Pi^a \partial^\mu \Pi^a
              + \frac{\Sigma m}{F^2} \Pi^a \Pi^a.
\ee
This results in the pion propagator $1/(p^2 +m_\pi^2)$ with pion masses
given by the Gellmann-Oakes-Renner relation
\be
m_\pi^2 =\frac{2m\Sigma}{F^2}.
\ee

In the long-wavelength limit the order of magnitude of the different
terms contributing
to the action of the chiral Lagrangian is given by \cite{GL}
\be
S = \int d^4 x {\cal L} (U) \sim L^{d-2} {\Pi^{a}_{NZM}}^2
      + L^d \frac {\Sigma m}{F^2} ({\Pi^{a}_{ZM}}^2+{\Pi^{a}_{NZM}}^2).
\ee
Here, the $\Pi^{a}_{ZM}$ represent the zero momentum modes with no space time 
dependence, whereas the nonzero momentum modes are denoted  by
by $\Pi^{a}_{NZM}(x)$. This decomposition has two immediate 
consequences. First,
for $\frac {\Sigma m}{F^2} \gg \frac 1V$ the fluctuations of the pion fields
are small and it is justified to expand $U$ in powers of $\Pi^a$. Second,
for
\be
\frac {\Sigma m}{F^2} \ll\frac 1{\sqrt V}
\ee
the fluctuations of the zero modes dominate the fluctuations of the
nonzero modes, and only the contribution from the zero modes 
has to be taken into account for the calculation of an observable.
In this limit the so called finite volume partition function is given by
\be
Z^{\rm eff}_{N_f}({\cal M},\theta)
\sim \int_{U\in SU(N_f)} dU e^{V\Sigma {\rm Re\,Tr}\,
{\cal M} Ue^{i\theta/N_f}},
\label{zeff}
\ee
where the $\theta$-dependence follows from the $\theta-$dependence of the QCD
partition function via the combination \cite{LS} $m e^{i\theta/N_f}$. 
We emphasize that any theory with the same pattern of chiral symmetry
breaking as QCD can be reduced to  the same extreme infrared limit.
\subsection{Leutwyler-Smilga Sum Rules}

The finite volume
partition function in the sector of topological charge $\nu$ follows
by Fourier inversion according to (\ref{ZQCDdet}).
The partition function for $\nu = 0$ is thus given by (\ref{zeff})
with the integration over $SU(N_f)$ replaced by an integral over
$U(N_f)$. 

The Leutwyler-Smilga sum-rules \cite{LS} are obtained 
by expanding the partition
function  in powers of $m$ before and after
averaging over the gauge field configurations and equating the
coefficients. This corresponds to an expansion in powers of $m$ of
both the QCD partition function (\ref{ZQCD}) and the finite
volume partition function (\ref{zeff}) in the sector
of topological charge $\nu$.
As an example,
we consider the coefficients of $m^2$ in the
sector with $\nu = 0$. After performing the group integrals we find
the sum-rule
\be
\langle {\sum}' \frac 1{\lambda_k^2}\rangle  = \frac {\Sigma^2 V^2}{4N_f},
\label{LS2}
\ee
where the prime indicates that the sum is restricted to nonzero positive
eigenvalues.

By equating higher powers of $m^2$ one can generate an infinite family
of sum-rules for the eigenvalues of the Dirac operator. However, they
are not sufficient to determine the Dirac spectrum. The reason is
that  the mass in the propagator also
occurs in the fermion-determinant of the QCD partition function. 
However, as will be shown
in the next section, the Dirac spectrum can be obtained from a chiral
Lagrangian corresponding to QCD with additional bosonic and fermionic 
quarks \cite{OTV}.
We conclude that chiral symmetry breaking leads to correlations
of the inverse eigenvalues which are determined by the underlying
global symmetries.

\subsection{The Partially Quenched QCD Partition Function}

As was mentioned in previous section, the valence quark mass dependence of
the chiral condensate cannot be extracted from the QCD partition function.
The solution to this problem is to simultaneously introduce yet
another quark species of opposite statistics
\cite{OTV}. 
This corresponds to a Euclidean partition function of the form

\be
Z^{\rm pq}  ~=~ 
\int\! dA
~\frac{\det(D + m_{v1})}{\det(D + m_{v2})}\prod_{f=1}^{N_{f}}
\det(D + m_f) ~e^{-S_{YM}} ~,
\label{pqQCD}
\ee
which we will call the partially quenched or pq-QCD partition function.
When $m_{v1}=m_{v2}$ this partition function simply coincides with the original
QCD partition function. 
However, it is now also the generator of a mass-dependent chiral
condensate (see (\ref{massdep}))
for the additional (say, fermionic) quark species. In the sector of topological
charge $\nu$ we find
\be
\Sigma(m_{v};m_1,\cdots, m_{N_f}) =
\frac 1V \left .\frac {\partial}{\partial m_{v1}}
\right |_{m_{v1} = m_{v2}= m_v} \log Z_\nu^{\rm pq},
\label{valrel}
\ee
where $Z_\nu^{\rm pq}$ is the partially quenched QCD partition function
in the sector of topological charge $\nu$.

Our aim is to find the chiral Lagrangian corresponding to (\ref{pqQCD}).
If we are successful, we have succeeded in deriving a partition function
for the extreme 
infrared limit of the spectrum of the QCD Dirac operator. These questions
will be addressed in the next sections.

\subsection{The Infrared Limit of QCD}

If we don't write the determinants in terms of integrals
over complex conjugated variables, the
global flavor symmetry of the partially quenched 
QCD partition function (\ref{pqQCD}) is broken spontaneously
according to
\be
Gl_R(N_f+1|1) \times Gl_L(N_f+1|1) \rightarrow Gl_V(N_f+1|1)
\label{susybreaking}
\ee
with an axial $U(1)$ group broken explicitly by the anomaly. 
Here, the groups $Gl_R(N_f+1|1)$,  $Gl_L(N_f+1|1)$ and $Gl_V(N_f+1|1)$
are super-groups of matrices acting on vectors with $N_f+1$ fermionic
components and one bosonic component. The subscript refers to
right-handed (R), left-handed (L) and vector (V), respectively. 
The latter transformations transform the right-handed and the left-handed
fermion fields in the same way.
For a confining
theory such as QCD the only low-lying modes are the Goldstone modes 
associated with the spontaneous breaking of chiral symmetry. The quark masses
play the role of symmetry breaking fields. 

Although the axial supergroup $Gl_A(N_f+1|1)$ 
is a symmetry group of the pq-QCD action 
(\ref{pqQCD}) it is not
necessarily a symmetry of the QCD partition function. It may be that the
symmetry transformations violate the convergence of the integrals in
 the partition function. There are no problems for the Grassmann variables.
However, the integrations over the bosonic fields are only convergent if
the fields are related by complex conjugation. For this reason a $U_A(1)$
transformation in which the bosonic fields
$\phi_R$ and $\phi_L$ are transformed according
to a different phase factor is not a symmetry of the partition
function. What is a symmetry of the partition function is the 
axial transformation
\be
\phi_L \rightarrow e^s \phi_L,\qquad \phi_L^* \rightarrow e^s \phi_L^*,\\
\phi_R^* \rightarrow e^{-s} \phi_R^* \qquad
\phi_R \rightarrow e^{-s} \phi_R.
\ee
Mathematically, this symmetry group is $Gl(1)/U(1)$. Had we restricted
ourselves to the unitary subgroup $U(N_f+1|1)$ of $Gl(N_f+1|1)$ 
from the start, we
would  have missed this class of symmetry
transformations.

Taking into account the chiral anomaly, the chiral symmetry in
(\ref{pqQCD}) is broken spontaneously according to
(\ref{susybreaking}).
The symmetry of the QCD partition function is thus reduced to
$Sl_V(N_f+1|1)\oslash Gl_V(1)$ where $\oslash$ denotes the semi-direct
product. 
The Goldstone manifold corresponding to the symmetry breaking pattern
(\ref{susybreaking}) is based on
the symmetric superspace $Sl_A(N_f+1|1)$. In our effective
partition function the terms that break the axial symmetry will be
included explicitly resulting in an integration manifold given by
$Sl_A(N_f+1|1) \otimes Gl_A(1)$. However, this manifold is
not a super-Riemannian manifold and is not suitable as an integration domain
for the low energy partition function. As an integration domain we choose the
 maximum Riemannian submanifold of $Gl_A(N_f+1|1)$.
This results in a  fermion-fermion
block given by the compact domain $U_A(N_f+1)$,
whereas the boson-boson
block is restricted to the non-compact domain $Gl(1)/U(1)$. Because of
the super-trace, this compact/non-compact structure is required for obtaining
a positive definite quadratic form for the mass term and the
kinetic term of our low energy effective partition function
\cite{ChPTfound}. For a detailed
mathematical discussion of this construction we refer to a paper by
Zirnbauer \cite{Zirnall}.

To lowest order in the momenta, the infrared limit of the QCD partition 
function is uniquely determined by the geometry of the Goldstone 
manifold and Lorentz invariance. With the 
singlet field $\Phi_0 \equiv {\rm Str \Phi}$ fluctuating about the vacuum
angle $\theta$ with an amplitude given by the singlet mass
we thus obtain the effective partition function
  \cite{Morel,pqChPT}
\be
Z(\theta,\hat {\cal M}) &=& \int_{U\in Gl(N_f+1|1)} dU \exp \int d^4 x \left [
\frac{F^2}{4} \;
{\rm Str} (\partial_\mu U \; \partial_\mu U^{-1})
\right .\nonumber \\
&+& \left . \frac{\Sigma}{2} \; {\rm Str} (\hat{\cal M} U+\hat{\cal M} U^{-1})
+\frac{F^2 m_0^2}{12} \; (\frac {\sqrt 2 \Phi_0}F-\theta)^2\right ].
\nonumber \\
\label{zthetam}
\ee
This partition function contains terms up to order $p^2$. The kinetic term
of the singlet field is subleading and has not been included.
The supertrace (Str) and the superdeterminant (Sdet) 
of a graded matrix with bosonic blocks $a$ and $b$ and 
fermionic blocks $\sigma$ and $\rho$ are defined as
\cite{Berezin,Efetovbook}
\be
{\rm Str} 
\left ( \begin{array}{cc} a &\sigma \\ \rho & b\end{array} \right )
= {\rm Tr} a - {\rm Tr } b,
\label{supertrace}
\ee
and
\be
{\rm Sdet}
= \frac {\det(a - \sigma b^{-1} \rho)}{\det b}.
\label{superdeterminant}
\ee
Recently, this partition function for $N_f = 0$ 
was derived from a two sub-lattice random flux model \cite{Simons-Altland}
using the flavor-color transformation introduced by 
Zirnbauer \cite{color-flavor}.
In our language,  such model is QCD at infinite coupling with $U(N)$ gauge
fields and quenched Kogut-Susskind fermions with no phase factor from the
$\gamma$ matrices. In this case, as well as in the effective partition
function for Kogut-Susskind fermion, the singlet mass term is absent. In both
cases the $U(1) \times U(1)$ symmetry  is broken 
spontaneously to $U(1)$ instead of broken explicitly by the anomaly. The reason
is that the $U(1)$ symmetry of Kogut-Susskind fermions is 
an axial isospin transformation rather than the  singlet $U_A(1)$ 
transformation (We thank Paul Rakow and Misha Stephanov for 
discussions to clarify this point.).

By a generalization of an argument due to Gasser and Leutwyler \cite{GL}
to be discussed in the next section
it can be shown that for masses in the range (\ref{thouless})
this partition function factorizes into a product over zero momentum modes
and non-zero momentum modes \cite{Osbornprl,OTV}.
To leading order, the mass dependence of the of the QCD partition
function is thus  given by
\be
Z_{\rm eff}(\theta,\hat {\cal M}) = \int_{U\in Gl(N_f+1|1)} dU e^{
V\frac{\Sigma}{2} \; {\rm Str} (\hat{\cal M} U+\hat{\cal M} U^{-1})
-\frac{F^2 m_0^2 V}{12} \; (\frac{\sqrt 2 \Phi_0}F-\theta)^2}.
\ee
The partition function in a sector of topological charge $\nu$ 
follows by Fourier inversion \cite{LS}
\be
Z^\nu_{\rm eff}(\hat{\cal M}) &=& 
\frac 1{2\pi} \int_0^{2\pi} d\theta  e^{-i\nu \theta}
Z_{\rm eff}(\theta,\hat {\cal M}).
\ee
We perform the integration over $\theta$  after a Hubbard-Stratonovitch
transformation which linearizes the singlet mass term.
Up to a mass independent factor this results in the
partition function \cite{OTV}
\be
Z^\nu_{\rm eff}(\hat{\cal M}) =
\int_{U\in Gl(N_f+1|1)} dU \,{\rm Sdet}^\nu U \, e^{
 V\frac{\Sigma}{2} \; {\rm Str} (\hat{\cal M} U+\hat{\cal
M} U^{-1})}.
\label{superpart}
\ee

     \subsection{Domains in (Partially-Quenched) Chiral Perturbation Theory} 

 In chiral perturbation theory the different domains of validity where
analyzed in detail by Gasser and Leutwyler \cite{GL}. A similar analysis
applies to partially quenched chiral perturbation theory. The idea is
as follows. The $U$ field can be decomposed as \cite{GL}
\be
U = U_0 e^{i\psi(x)}.
\ee
where $U_0$ is a constant (zero-momentum) field. The kinetic term of the
low-momentum components is the  $\psi$ fields  can be approximated by
\be
\frac 12\int d^4 x \partial_\mu \psi^a \partial_\mu \psi^a  \sim L^2 \psi^2.
\ee
We observe that the magnitude of the fluctuations of the $\psi$ field
are of order  $1/L$ which justifies a perturbative expansion of
$\exp(i \psi(x))$. The fluctuations of the zero modes (i.e. constant fields), 
on the other hand, are only limited by the mass term
\be
\frac 12  V \Sigma {\rm Str} {\cal M}(U_0+ U_0^{-1}).
\ee
For quark masses 
$m \gg 1/V\Sigma$, the field $U_0$ fluctuates close to the identity and
the $U_0$ field can be expanded around the identity as well. 
This is the  domain of chiral perturbation theory.

 For valence quark masses in the range
\be
\frac 1{V\Sigma}\ll  m_v \ll \frac{F^2}{\Sigma L^2}
\ee
the valence quark mass dependence of the chiral condensate is dominated by
the zero momentum modes
which  can be treated perturbatively. Below we will show that in this
domain chiral perturbation theory and random matrix theory coincide.
In the theory of disordered mesoscopic systems it is well-known that
random matrix theory is valid below an energy scale
$\sim 1/L^2$. This will be 
discussed in the next section.

     \subsection{Picture from Mesoscopic Physics}

In disordered mesoscopic physics it has been found that Random Matrix
Theory is valid for eigenvalues separated less than the energy scale 
$E_c$ (see eq. (\ref{thoulessm})).
In this context
$E_c$ is defined as the inverse tunneling time of an electron through
the sample. For diffusive
motion, the distance $\Delta x$ a particle diffuses in time interval
$\Delta \tau$ is given by 
\be
(\Delta x)^2 = D \Delta \tau,
\ee
where $D$ is the diffusion constant.
The diffusion time through the sample is thus given by $L^2/D$ resulting
in a Thouless energy of
\be
E_c = \frac{\hbar D}{L^2}.
\ee

A second energy scale that enters in  mesoscopic physics is $\hbar/\tau_e$,
where $\tau_e$ is the elastic scattering time.
Based one these two scales three different domains for
the energy difference, $\delta E$, that enters in the two-point
correlation function, can be distinguished:
the ergodic domain, the diffusive domain
or Altshuler-Shklovskii domain and the ballistic
domain.
For energy differences $\delta E \ll E_c$ eigenvalue
correlations are given by RMT.
Since for time scales larger than the diffusion time, an initially localized
wave packet explores the complete phase space,
this regime is known as the ergodic regime. In the diffusive
regime defined by
 $E_c \ll \delta E\ll \hbar/\tau_e$
only  part of the phase space
is explored by an initially localized  wave packet,
resulting in the disappearance of eigenvalue correlations.
In this paper we don't consider the ballistic regime with $\delta E \gg
\hbar/\tau_e$.
For an interpretation of the Thouless
energy in terms of the spreading width we refer to \cite{spreading,Guhr}.
As was shown in \cite{Imry,kravtsovmoriond},
the spectral two-point function can be related to
the semiclassical return probability which provides a simple intuitive
picture of  its asymptotic behavior.
For other recent studies on this topic we refer to
\cite{Alt,Braun,Aronov,yan,kravtsov-lerner,Altland-Gefen,Chalker-kravtsov}.
What has emerged from these studies is that there
is a close relation between eigenvalue correlations and localization
properties of the wave functions.

\begin{figure}[!ht]
\centering
\includegraphics[width=75mm, height=55mm]{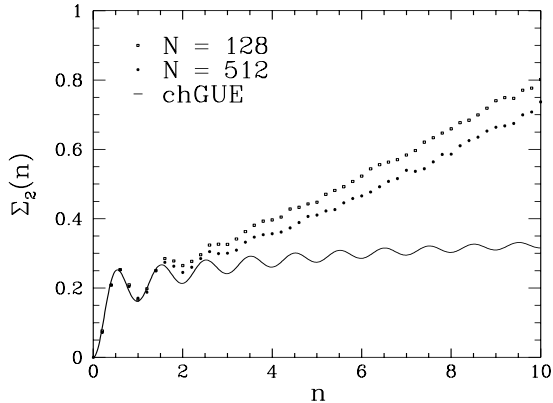}
\caption{The number variance $\Sigma_2(n)$ versus $n$
approximation for an interval starting at $\lambda = 0$. The total number
of instantons is denoted by $N$.}
\end{figure}

Based on these ideas we can interpret the Dirac spectrum as the energy levels
of a system in 4 Euclidean dimensions and one artificial time dimension.
The corresponding classical evolution can be easily derived from the
Heisenberg equations of motion
\be
\frac {dx}{d\tau} = i[x, D].
\ee
According to the Bohigas conjecture the eigenvalue correlations are given by
RMT if and only if the corresponding classical motion chaotic. We thus
conclude  that the classical time evolution of quarks in the Yang-Mills
gauge fields is chaotic.

These ideas can be tested by means of lattice QCD \cite{many,Guhr-Wilke}
 and instanton liquid~\cite{Osbornprl}
simulations. In Fig. 1, we show \cite{Osbornprl} the number variance
$\Sigma_2(n)$,
defined as the variance of
the number of levels in an interval containing $n$ level on average, versus
$n$ for eigenvalues obtained
from the Dirac operator in the background of instanton liquid gauge
field configurations.
The chRMT result,  given by the solid curve, is reproduced
up to about two level spacings.
In units of the  average level spacing, $\Delta=1/\rho(0)=\pi/\Sigma V$,
the energy $E_c$ is given by
$n_c \equiv {E_c}/{\Delta} = {F^2 L^2}/\pi$.
For an instanton liquid with instanton density $N/V =1$ we find that
$n_c \approx 0.07 \sqrt N$. We conclude that chRMT appears to
describe the eigenvalue correlations up to the
predicted scale.

In mesoscopic physics the Goldstone modes that enter in the 
theory of impurity scattering are known as the diffusons. The reason
is that the Goldstone propagator satisfies a diffusion equation 
(see for example the book by Efetov \cite{Efetovbook}). It describes
the  diffusion of electrons by impurity
scattering \cite{Stone}. Similarly,  the pion propagator also satisfies 
a diffusion equation. This can be seen  by Fourier transforming
the propagator with regards to the imaginary valence masses
$m_v = i \lambda$ and $m_{v'} = i \lambda'$,
\be
\Pi(x,\tau) = \int \frac {d^4p}{(2\pi)^4} \int d \lambda d \lambda'
\frac {e^{ipx + i\tau(\lambda+\lambda')}}{p^2 +(i\lambda 
+i\lambda')\Sigma/F^2}.
\ee
For $(x,\tau)\ne 0$ this propagator satisfies the diffusion equation
\be
(D\partial_x^2 - \partial_\tau)\Pi(x,\tau) = 0
\ee
with a diffusion constant given by \cite{Osbornprl,janikdif} 
$D= F^2/\Sigma$ .

\section{Calculation of Valence Quark Mass Dependence of the Chiral 
Condensate from QCD}

\subsection{Nonperturbative Evaluation of $\Sigma(m_v)$}

In this section we calculate the valence quark mass dependence of the
chiral condensate for the simplest case of $N_f = 0$ and $\nu =0$ in the
domain $m_v \ll F^2/\Sigma L^2$. In this domain
the partition function is given by
\be
Z(J) = \int_{U\in Gl(1|1)} dU \exp\left [ \frac {\Sigma V}2 {\rm Str}
\left ( \begin{array}{cc} m_v +J &0 \\ 0 & m_v \end{array} \right )
(U + U^{-1})\right ],
\ee
where the integration is over the maximum super-Riemannian sub-manifold
of $Gl(1|1)$. This manifold is parametrized by
\be
U =\exp {\left ( \begin{array}{cc} 0 &\alpha \\ \beta & 0 \end{array} \right )}
 \left ( \begin{array}{cc} e^{i\phi} &0 \\ 0 & e^s \end{array} \right ).
\ee
The integration measure is the Haar measure which in 
terms of this parametrization
it is given by
\be
{\rm Sdet } \frac {\delta U_{kl}}{
\delta \phi \, \delta s\, \delta \alpha \, \delta \beta} 
\, \,d\alpha d\beta d\phi ds,
\ee
where $\delta U \equiv U^{-1} dU$.

It is straightforward to calculate the Berezinian going from the
variables $\delta U_{11}\,\delta U_{22}\,
\delta U_{12}\,\delta U_{21}$ to the variables 
$\delta \phi \, \delta s\, \delta \alpha \, \delta \beta$. The 
derivative matrix is given by
\be
B=\frac {\delta U_{kl}}{
\delta \phi \, \delta s\, \delta \alpha \, \delta \beta} =
\left (\begin{array}{cccc} 
i & 0 &  \beta /2&
 \alpha/ 2\\
0  &1 &  \beta/ 2 &\alpha /2\\
0 & 0& e^{s-i\phi}& 0 \\
0 & 0 & 0 &e^{-s+i\phi} 
\end{array}\right).
\ee
Using the definition of the graded determinant one simply finds that
${\rm Sdet} B = i$. Up to a constant, the integration measure is thus given by
$d\phi ds d\alpha d\beta$. In general, for $N_f \ne  0$, the Berezinian
is more complicated \cite{Berezin,DOTV}.

We also need
\be
\frac 12(U +U^{-1} ) = \left ( \begin{array}{cc} 
 (1 +\frac {\alpha \beta} 2 ) \cos \phi& \alpha (e^s - e^{-i\phi})\\
\beta(e^{i\phi} - e^{-s} ) & \cosh s ( 1- \frac{\alpha \beta }2) 
\end{array} \right ).
\ee
After differentiating with respect to the source term  
$(\Sigma(m_v) = \partial_J\log Z(J) |_{J=0}/V)$ this results in
\be
\frac{\Sigma (m_v)}{\Sigma} =  \int \frac{d\phi ds d\alpha d\beta}{2\pi}
\cos \phi (1 +\frac {\alpha \beta} 2 ) 
e^{x \cos \phi (1 +\frac {\alpha \beta} 2 ) -
x\cosh s ( 1- \frac{\alpha \beta }2)}.\nonumber \\
\ee
With the Grassmann integral given by the coefficient of $\alpha\beta$ 
we  obtain
\be
\frac{\Sigma(m_v)}{\Sigma} &=& \int \frac{ds d\phi}{4\pi} 
[\cos\phi  \nonumber \\&+&x(\cos\phi 
+ \cosh s)\cos\phi ]e^{x(\cos\phi - \cosh s )} .
\ee
Now all integrals can be expressed in terms of modified Bessel functions.
We find
\be
\frac{\Sigma(m_v)}{\Sigma} &=& I_1(x)K_0(x)
+\frac x2 ( I_2(x)K_0(x)+I_0(x)K_0(x)+ 2I_1(x)K_1(x)).
\ee
After using the recursion relation for  modified Bessel functions,
\be
I_2(x) = I_0(x) -\frac 2x I_1(x),
\ee
 we arrive at the final result \cite{vPLB,OTV,DOTV}
\be
\frac{\Sigma(m_v)}{\Sigma} = x(I_0(x)K_0(x)+ I_1(x)K_1(x)),
\ee
where $x= mV\Sigma$.

This calculation can be generalized to arbitrary $N_f$ and arbitrary $\nu$. 
The calculation for arbitrary $N_f$ is much more complicated, but with a
natural generalization of the factorized parametrization, and using some known
integrals over the unitary group \cite{brower}, 
one arrives at an
expression in terms of modified Bessel functions
\be
\frac {\Sigma(x)}{\Sigma} = x(I_{a}(x)K_{a}(x)
+I_{a+1}(x)K_{a-1}(x)),
\label{val}
\ee
where $a = N_f+|\nu|$. This result is in 
complete agreement \cite{DOTV} with chRMT to be discussed below.

\subsection{Perturbative Calculation}

The valence quark mass dependence of the chiral condensate follows
from (see eq. (\ref{valrel}))
\be
\Sigma(m_v)= \frac 1V
\left . \partial_{J} \log Z_{\rm eff}^{\nu}(\hat {\cal M}) \right |_{J=0}.
\ee
In the quenched limit this results in (see eq. (\ref{zthetam}) for
the definition $S(U(x))$)
\be
\Sigma(m_v)= \int_{U(x)\in Gl(1|1)} dU(x) \frac{\Sigma V}2 
\partial_J {\rm Str}
{\cal M}
(U(x) + U^{-1}(x)) e^{S(U(x))},
\ee
Perturbatively to one loop order we can expand $U$ around the identity, and
do the usual Gaussian integrations for a flat measure. The trace of the
propagator
of the Goldstone bosons containing valence quarks is given by
\be
\langle \pi_k^2 \rangle = \frac 1V \sum_p \frac {(-1)^L}{p^2 +M_{\pi}^2},
\ee
where $L$ is 1 for a bosonic pion and $1$ for a fermionic pion.
The valence pion mass, $M_\pi$, is given by the 
Gell-Mann-Oakes-Renner relation 
\be
M_{{vv}} = \frac {2 m_v \Sigma}{F^2}, \quad \hbox{or}
\quad M_{{vs}} = \frac { (m_v+m_s) \Sigma}{F^2}, \quad \hbox{or}
\quad M_{{ss}} = \frac { 2m_s \Sigma}{F^2}.
\ee

The calculation of the valence quark-mass dependence of the chiral condensate
using the partially quenched chiral Lagrangian is basically straightforward. 
The only complication is the "$\eta'$"-mass term (the terms proportional to
$m_0^2$ in the Lagrangian). However, the quadratic form in the pion fields
can be diagonalized \cite{pqChPT}
resulting in an analytical expression for the 
propagator. In the limit that the $\eta'$ mass is much larger than the mass
of the Goldstone bosons, we find \cite{GolLeung,OTV}
\be
\frac {\Sigma_v(m_v)}{\Sigma}= 
1-\frac{1}{N_f F^2} \Bigg\{ N^2_f \Delta(M_{sv}^2)- \Delta(M_{vv}^2)
 + (M^2_{ss}-M^2_{vv}) \; \partial_{M_{vv}^2}
\Delta(M_{vv}^2) \Bigg\}  .\nonumber\\
\label{generalsigma}
\ee
where 
the trace of the finite volume pion propagator is given by
\be
\Delta(M^2)= \frac  1V \sum_p   \;
\frac{1}{p^2+M^2}.
\ee
Here, the sum is over momenta in a box of volume $L^4$. From a careful analysis
of the diverging momentum summations it follows that
this propagator can be expanded in powers of $M^2$ 
\be
\Delta(M^2) = \frac 1{M^2L^4} - \frac {\beta_1}{L^2} + O(\frac{M^2}
{\Lambda^2}),
\label{expandprop}
\ee
where $\beta_1 = 0.140461$ for a 4-dimensional hyper-cubic box, and $\Lambda$
is the momentum cutoff \cite{HasLeu}. In agreement
with a naive inspection of the momentum sum, the propagator is dominated
by the zero momentum term if
\be
              M \ll \frac 1L,
\ee
or the Compton wavelength of the valence pion is much larger than the size
of the box. As discussed in section (3.1) 
we thus find that for
\be
m_v \ll \frac{F^2}{\Sigma L^2}
\ee
the valence quark mass dependence of the chiral condensate is described
by the zero momentum contribution to the trace of the propagator. 
Let us investigate
the zero momentum limit of (\ref{generalsigma}) given by replacing
the propagator by the first term in (\ref{expandprop}). In the chiral
limit with $m_s \rightarrow 0$ we then find
\be
\Sigma_v &\sim& \Sigma(1 - \frac {N_f} {V M_{sv}^2 F^2})\nonumber \\
         &\sim &  \Sigma(1 - \frac {N_f} { m_vV \Sigma}).
\label{intermediate}
\ee
The nonperturbative result for $\Sigma(m_v)$ obtained
for a fixed
value of the topological charge (\ref{val})
$\nu$ instead of at $\theta = 0$ can
also be expanded  in powers of 
$1/x\equiv 1/m_vV\Sigma$
\be
\frac {\Sigma(x)}{\Sigma} \sim 1- \frac {N_f+|\nu|}{x} +
\frac {|\nu|} x +O(\frac 1{x^3}),
\ee
where the  term $|\nu|/x$ results from the factor $m^{|\nu|}$ in the partition
function \cite{msumrules}. 
We find that to leading order in $1/x$ the valence quark mass
dependence of the chiral condensate does not depend on the topological
charge and agrees with  chiral perturbation theory.

The result for the valence quark mass dependence of the chiral condensate
(\ref{generalsigma}) is valid in the diffusive domain as well. In that
case it is possible to rederive and to extend results by Smilga and
Stern \cite{Smilga-Stern} for the slope of the Dirac spectrum.

      \subsection{Two-Point Correlation Function}
Let us do a perturbative calculation of the two-point function in the
domain that is dominated by the zero-momentum modes, i.e. for valence
quark masses well below $F^2/\Sigma L^2$. The generating function
is defined by
\be Z(J,J') =\int_{U\in Gl(2|2)} dU e^{S(U)},
\ee
where the action is given by
\be
S(U) = \frac {\Sigma V}2 {\rm Str} 
\left (\begin{array}{cccc} m+J & 0 & 0& 0 \\
                            0 & m'+J' & 0& 0 \\
                          0 & 0 & m& 0 \\
                          0 & 0 & 0& m'     \end{array} \right )
         (U+U^{-1}).
\ee
The spectral two-point  correlation function follows from the
partition function by \cite{OTV}
\be
\langle \rho(\lambda) \rho(\lambda') \rangle =
\frac 1{4\pi^2} {\rm Disc} \partial_J \partial_{J'} \left . Z(J,J') 
\right |_{J=0, J' = 0, m =i\lambda,m' = i\lambda'}.
\ee 

We are interested in the discontinuity of the  connected part of the
correlation function. The only contribution is from mesons containing
a valence quark with mass $m$ and a valence quark with mass $m'$. If 
we expand $U \equiv \exp i \sqrt 2 \Pi_a T^a/F$ in powers of
$\Pi_a T^a/F$ we find that the connected spectral two-point function
\be
\langle \rho(\lambda) \rho(\lambda') \rangle_c = 
\frac {\Sigma^2}{F^4 4\pi^2} \left . {\rm Disc} \right |_{m =i\lambda, 
m' =i\lambda'} \frac 1{M_{vv'}^2},
\ee
has been expressed in terms of  the discontinuity 
of the valence pion susceptibility.
To obtain this result we have used among others that 
\be
{\rm Str} \left ( \begin{array}{cccc} 1 & 0 & 0& 0 \\
                            0 & 0 & 0& 0 \\
                          0 & 0 & 0& 0 \\
                          0 & 0 & 0& 0     \end{array} \right )T_a T^a = \frac 
12,
\ee
and the super-symmetry of the partition function.
To calculate the discontinuity we wish to remind the reader that 
\be
M_{vv'} = \frac {(|m| + |m'|) \Sigma}{F^2},
\ee
where $|m| = \sqrt {(m^2)}$. We finally find
\be
\langle \rho(\lambda) \rho(\lambda') \rangle_c = 
\frac {1}{2\pi^2} \left ( \frac 1{(\lambda +\lambda')^2} +
\frac 1{(\lambda -\lambda')^2}\right ),
\ee
which agrees with the asymptotic result for the spectral correlation
function \cite{VZ} for the chiral Random Matrix Theory to be discussed next.

\section{Chiral Random Matrix Theory}
     \subsection{Definition of the Model}
The chiral random matrix partition function
 with the global symmetries of the QCD partition function as
input is defined by {\cite{SVR,V}}
\be
Z^\nu_\beta({\cal M}) =
\int DW \prod_{f= 1}^{N_f} \det({\rm \cal D} +m_f)
e^{-\frac{N \beta}4 \Sigma{\rm Tr}W^\dagger W },
\label{ranpart}
\ee
where
\be
{\cal D} = \left (\begin{array}{cc} 0 & iW\\
iW^\dagger & 0 \end{array} \right ),
\label{diracop}
\ee
and $W$ is a $n\times m$ matrix with $\nu = |n-m|$ and
$N= n+m$.
As is the case in QCD, we assume that the equivalent of the topological charge
$\nu$ does not exceed $\sqrt N$,
so that, to a good approximation, $n = N/2$.
Then the parameter $\Sigma$ can be identified as the chiral condensate and
$N$ as the dimensionless volume of space time (Our units are defined
such that the density of the modes $N/V=1$).
The matrix elements of $W$ are either real ($\beta = 1$, chiral
Gaussian Orthogonal Ensemble (chGOE)), complex
($\beta = 2$, chiral Gaussian Unitary Ensemble (chGUE)),
or quaternion real ($\beta = 4$, chiral Gaussian Symplectic Ensemble (chGSE)).
For QCD with three or more colors and quarks in the fundamental representation
the matrix elements of the Dirac operator are complex and we have $\beta = 2$.
For $N_c=2$ and quarks in the fundamental representation the situation
is more interesting. As discussed in section (2.4) it is always possible 
to find
a basis for which the matrix elements of the continuum Dirac operator
are real for all gauge fields. Then the Dyson index of the corresponding chRMT
 is $\beta = 1$.
For Kogut-Susskind fermions in this case
the anti-unitary symmetry is different (see section (2.4)) 
and the matrix elements
of the Dirac operator can be organized into real quaternions  corresponding
to a chRMT with $\beta = 4$. For gauge fields in the adjoint
representation the gauge fields are real resulting in an anti-unitary 
symmetry also corresponding to the class $\beta = 4$ for any value
of $N_c$.

The reason for choosing a Gaussian distribution of the matrix elements is
its mathematically simplicity. This model can be generalized to arbitrary
potential
\be
{\rm Tr}W^\dagger W \rightarrow {\rm Tr} V(W^\dagger W )
\ee
where $V(x)$ is such that the resulting probability distribution
is well defined. In section (5.4) we will argue that the interesting 
properties of chRMT do not depend on the choice of $V(x)$. 

Below we will also discuss the invariant or Dyson-Wigner Random Matrix
Ensembles. 
They are defined as ensembles of Hermitean
matrices $\{H\}$
with independently Gaussian distributed matrix elements, i.e.
with probability distribution given by
\be
P(H) \sim e^{-\frac {N\beta} 2 {\rm Tr } H^\dagger H}.
\label{phinv}
\ee
Depending on the anti-unitary symmetry, the matrix elements are real, complex
or quaternion real. They are called the Gaussian Orthogonal Ensemble (GOE),
the Gaussian Unitary Ensemble (GUE) and the Gaussian Symplectic Ensemble
(GSE), respectively. Each ensemble is characterized by its Dyson index
$\beta$ which is defined as the number of independent variables per matrix
element. For the GOE, GUE and the GSE we thus have $\beta =1, \, 2$ and 4,
respectively.

Together with the Wigner-Dyson ensembles and 
the superconducting random matrix ensembles
\cite{Altland}
the chiral ensembles can be classified according to the Cartan classification
of large symmetric spaces \cite{Zirnall}.

 \subsection{Symmetries}
From the structure of the determinant, it follows immediately that 
the chiral random matrix partition function has the same global flavor
and $U_A(1)$ symmetries as the QCD partition function. 

It can be shown that in the domain (\ref{thouless}) the random matrix
partition function can be mapped onto the effective
finite volume partition function
\cite{SVR}. We will discuss such derivation in section (7.3) where 
we extend this model to finite temperature. 
The most detailed studies to this model were performed in the 
quenched limit by means of the supersymmetric method
\cite{MIT,Sener1,GWu,Seneru}. In that case the chRMT partition function
can be mapped onto a super-symmetric nonlinear $\sigma$-model.

In this model chiral symmetry is broken spontaneously with chiral condensate
given by  $\Sigma = 
\lim_{N\rightarrow \infty} {\pi \rho(\lambda\rightarrow 0)}/N$, where
$N$ is interpreted as the (dimensionless) volume of space
time. For complex matrix elements ($\beta =2$), which is appropriate for QCD
with three or more colors and fundamental fermions, the
symmetry breaking pattern is {\cite{SmV}} $SU(N_f) \times SU(N_f)/SU(N_f)$.
For $\beta = 1$ and 4 the symmetry breaking pattern is 
$SU(2N_f)/Sp(N_f)$ and $SU(N_f)/O(N_f)$ 
respectively \cite{SmV}, the same as in QCD {\cite{Shifman-three}}. 

Finally, one of the reasons for the mathematical simplicity of this model
is that it has more symmetry than the QCD partition function
The chRMT partition function is invariant with respect to the 
unitary transformation
\be
W \rightarrow U W V^{-1}.
\ee
Since an arbitrary complex matrix can be decomposed as
\be
W= U \Lambda V^{-1},
\label{polar}
\ee
with $\Lambda$ a diagonal matrix and $U$ and $V$ unitary matrices, the
chRMT partition function for arbitrary potential $V(\lambda)$ 
can be expressed in terms of the 
eigenvalues $\lambda_k$ as
\be
Z^{\nu}_{\beta}({\cal M}) =
\int d\lambda |\Delta(\lambda_k^2)|^\beta
\prod_k \lambda_k^{\beta\nu+\beta-1} \prod_f m_f^\nu
\prod_{f,k}(\lambda_k^2 + m^2_f)
e^{-\frac{N\beta}4 \sum_k V(\lambda_k^2)},\nonumber\\
\label{zeig}
\ee
where the Vandermonde determinant is defined by
\be
\Delta(\lambda_k^2) = \prod_{k<l} (\lambda_k^2-\lambda_l^2).
\label{vandermonde}
\ee
This result greatly simplifies the
mathematical treatment of the chRMT partition function. For example,
it allows us to use the orthogonal polynomial method.
We wish to point out that for $\beta=2$ and $m_f =0$ the ratio of the
partition function and $m^{N_f |\nu}|$ depends only on $\nu $ through 
the combination $N_f +|\nu|$. This duality between flavor and topology can
be understood more directly via Itzykson-Zuber integrals
\cite{camreview,wadati}.
\subsection{Properties of the Random Matrix Model}

The average spectral density that can be derived from (\ref{ranpart})
has the familiar semi-circular shape. As can be easily derived by means of the
orthogonal polynomial method, the
microscopic spectral density for the chGUE
is given by {\cite{SLEVIN-NAGAO,VZ,V}}
\be
\rho_S(u) = \frac u2 \left ( J^2_{a}(u) -
J_{a+1}(u)J_{a-1}(u)\right),
\label{micro2}
\ee
where $a = N_f + |\nu|$.
The valence quark mass dependence
of the chiral condensate follows from integration 
over the microscopic spectral density. If we use the microscopic variable
$u = \lambda V \Sigma$ as new 
integration variable, equation (\ref{massdep}) can be rewritten as
\be
\frac {\Sigma(m_v)}{\Sigma} = \int_0^\infty du \frac 
{2u \,du}{u^2 + (m_vV\Sigma)^2}
\frac 1{V\Sigma} \rho(\frac u{V\Sigma}).
\ee
In the thermodynamic limit,  the spectral density can be replaced
by the microscopic spectral density (\ref{micro2}). The integrals over
the Bessel functions are known and result in the following expression
for the valence quark mass dependence of the chiral condensate
\cite{vPLB}
\be
\frac {\Sigma(x)}{\Sigma} = x(I_{a}(x)K_{a}(x)
+I_{a+1}(x)K_{a-1}(x)),
\label{val2}
\ee
where $a = N_f+|\nu|$, $x = m_v V \Sigma$  
and $I_a$ and $K_a$ are modified Bessel functions. This result is in perfect
agreement with the result obtained from the partially quenched chiral
Lagrangian given in eq. (\ref{val}).
The spectral correlations near $\lambda=0$ can also be expressed in terms
of Bessel functions \cite{VZ}, whereas in the bulk of the spectrum
they are given 
by the invariant random matrix ensembles \cite{Kahn,nagao}. 
The microscopic spectral density and the microscopic spectral correlations
can also be derived for $\beta=1$ and $\beta=4$ but these two cases are
mathematically much more complicated. 
Many other
properties of the chRMT partition function have been calculated. Among others, 
we mention the distribution
of the smallest eigenvalue \cite{Forrester,Tilodam}
and the microscopic spectral density in the 
presence of nonzero quark masses \cite{Tilodam,DN,WGW,Jurk}.
For more discussion of the chRMT partition function we refer
to \cite{camreview}.

     \subsection{Universality}

The aim of universality studies is to identify observables that are
stable against deformations of the
random matrix ensemble. Not all observables have the same degree
of universality. For example, a semicircular average spectral density
is found for random matrix ensembles with independently distributed
matrix elements with a finite variance. However, this
spectral shape does not occur in nature, and it is thus not
surprising that it is only found in a rather narrow class of random matrix
ensembles. What is surprising is that the $microscopic$ spectral density
and the $microscopic$ spectral correlators are stable with respect to
a much larger class of deformations 
\cite{Bowick,Hack,Brezin,ADMN,KF,kanzieper,PZinnjustin,Eynard}
\cite{Brezin-Hikami,Beenakker}.
Two different types of deformations have been considered, those that
maintain the unitary invariance of the partition functions and those
that break the unitary invariance.

In the  first class, the Gaussian probability
distribution is replaced by $
P(W) \sim \exp(-N \sum_{k=1}^\infty a_k {\rm Tr}
(W^\dagger W)^k).$
For a potential with only $a_1$ and $a_2$ different from zero it was
shown \cite{Brezin} that
the microscopic spectral density is independent of $a_2$. A general proof
valid for arbitrary potential and all correlation functions
was given by Akemann et al. \cite{ADMN}.
The essence of the proof
is a remarkable generalization of the identity for
the Laguerre polynomials, $\lim_{n \rightarrow \infty}  L_n(x/ n) =
 J_0(2 \sqrt x) \ , $
to orthogonal polynomials determined by an arbitrary potential.
It was proved by taking the continuum limit of the recursion relation
for orthogonal polynomials.

In  the second class, an arbritrary fixed matrix is added
to $W$ in the Dirac operator (\ref{diracop}).
It has been shown that the microscopic spectral density and
the microscopic spectral correlations remain unaffected 
\cite{Sener1,GWu,Seneru}
for parameter values that completely modify the average spectral density.

Microscopic universality for deformations that affect the
macroscopic spectral density implies the existence of a scale
beyond which universality breaks down.
It can be interpreted naturally in terms of the
spreading width \cite{spreading} which  is the equivalent of the
Thouless energy.

Based on the general form of of the pqChPT partition function one
could argue that universality of the microscopic correlators in chRMT
is automatic.
However, one really has to show the stability of
the effective partition function with respect to variations of the
distribution of matrix elements. For the Wigner-Dyson ensembles the
stability of the saddle-point manifold was demonstrated in \cite{Hack}.

For reasons of mathematical simplicity, most  universality proofs have 
been performed for the $\beta=2$ ensembles. 
Recently, we have attempted to 
prove universality by establishing relations between the kernels for
the $\beta=1 $ and $\beta=4$ correlation functions and the kernel for
the $\beta = 2$ correlation functions. For the Gaussian ensembles such
relations are exact identities \cite{V2}. 
However, for an arbitrary potential
they are only valid asymptotically \cite{Senerprl,Widom}.

There are many other results related to the universality of chRMT. 
We mention recent results in QCD in three dimensions 
\cite{ADMN,DN3,Christiansen},
results for Dirac operators satisfying the Ginsparg-Wilson relation
\cite{Splittorf}, relations between the microscopic spectral density
and partition functions with two additional flavors
\cite{Dampart}.

\section{Chiral RMT as an  Exact Theory for the Fluctuations of Dirac 
         Eigenvalues}
     \subsection{Statistical Analysis of Spectra}
Spectra for a wide range of complex quantum systems
have been studied both experimentally and numerically (a excellent
recent review was given by Guhr, M\"uller-Groeling and
Weidenm\"uller {\cite{HDgang}}).
One basic observation
is that the scale of variations of the average spectral
density and the scale of the spectral fluctuations separate.
This allows us to unfold the spectrum, i.e. we rescale the
spectrum in units of the local average level spacing.
Specifically, the unfolded spectrum is given by
\be
\lambda_k^{\rm unf} = \int_{-\infty}^{\lambda_k} \langle \rho(\lambda')\rangle
d \lambda',
\ee
with unfolded spectral density
\be
\rho_{\rm unf} (\lambda) = \sum_k \delta(\lambda -\lambda_k^{\rm unf}).
\ee

The fluctuations of the
unfolded spectrum can be measured by suitable statistics. We will consider the
nearest neighbor spacing distribution, $P(S)$, and moments of the number of   
levels in an interval containing $n$ levels on average. In particular, we
will consider the
number variance, $\Sigma_2(n)$, and the first two cumulants, $\gamma_1(n)$ and
$\gamma_2(n)$. Another useful statistic is the
$\Delta_3(n)$-statistic introduced by Dyson and Mehta {\cite{deltaDM}}.
It is related to $\Sigma_2(n)$
via a smoothening kernel. The advantage of this statistic is that its
fluctuations as a function of $n$ are greatly reduced.
Both $\Sigma_2(n)$ and $\Delta_3(n)$ can be obtained from the pair correlation
function.

Analytical results for all spectral correlation functions
have been derived for each of the three ensembles {\cite{Mehta}}   
via the orthogonal polynomial method.
We only quote the most important results.
The nearest neighbor spacing distribution, which is known exactly in terms of a
power series, is well approximated by
\be
P(S) \sim S^{\beta}\exp(-a_\beta S^2),
\ee
where $a_\beta$ is a constant of order one.
The asymptotic behavior of $\Sigma_2(n)$
and $\Delta_3(n)$ is given by
\be
\Sigma_2(n) \sim (2/\pi^2\beta) \log n \quad {\rm and} \quad
\Delta_3(n) \sim \beta \Sigma_2(n)/2.
\ee
Characteristic features of random matrix correlations are
level repulsion at short distances and a strong suppression
of fluctuations at large distances.

For uncorrelated eigenvalues the level repulsion is absent and
one finds
\be
P(S) = \exp(-S),
\ee
and
\be
\Sigma_2(n) = n  \quad {\rm and} \quad \Delta_3(n) = n/15. 
\ee

\subsection{Results for Spectral Correlation Functions}

Recently, lattice QCD Dirac spectra were calculated and analyzed 
by a number of different groups. It was found that 
spectral correlations below a given scale are given by chRMT, whereas
above this scale, deviations from chRMT due to 
the nonzero momentum excitations were observed.
\cite{Halasz,vPLB,Trento,Tilo,Ma,Tilomore,many,Guhr-Wilke,Tilomass}
\cite{tilomark,tilochem,markum,Dam3d,DamSU3,TiloSU3,Tilo99,Berg} 
\cite{SVR,Vinst,Osbornprl,Osborn}. 

\begin{figure}[!ht]
\vbox{
\hbox{
\psfig{file=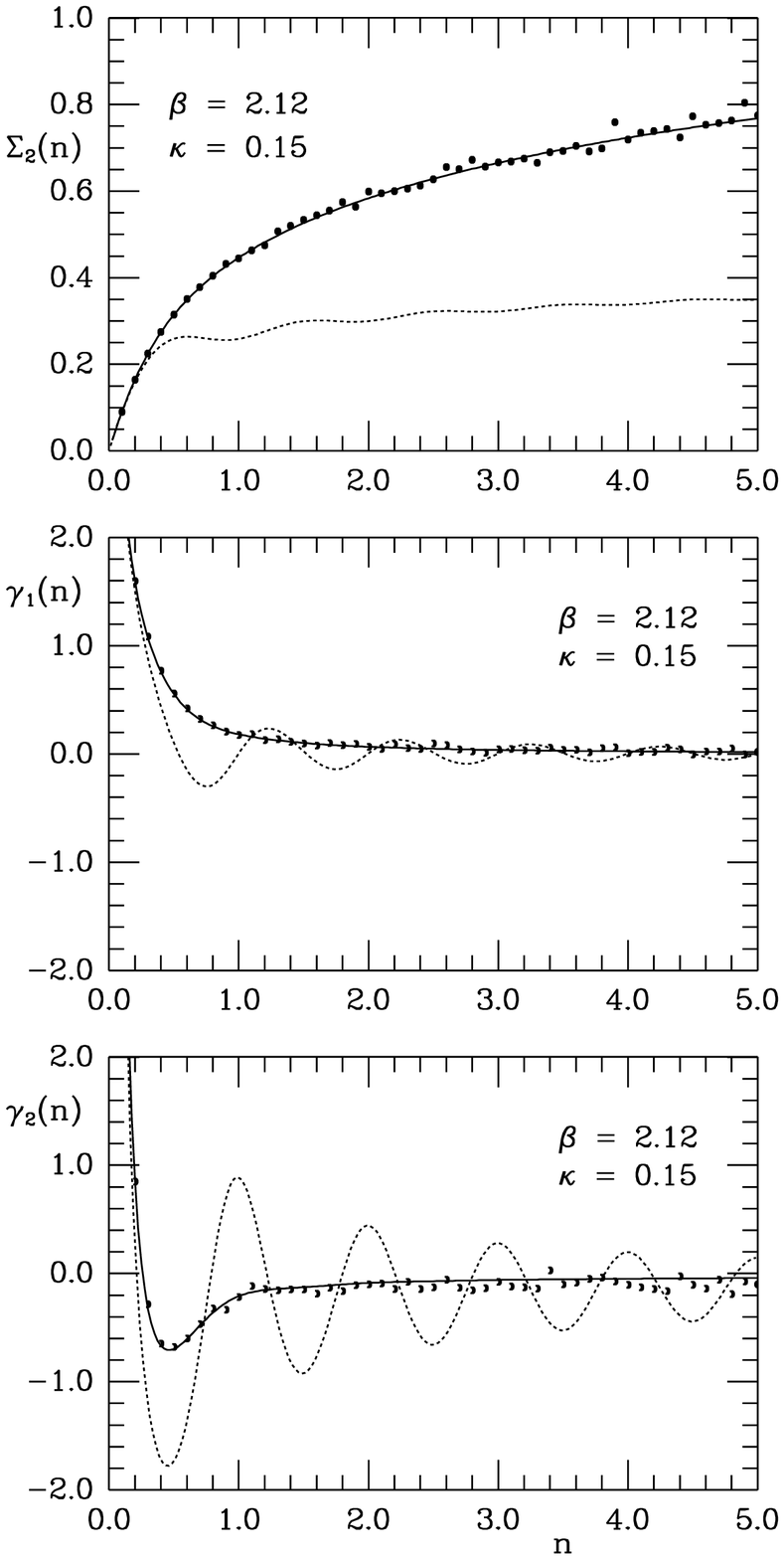,width=55mm}
\hspace*{1cm}
\psfig{file=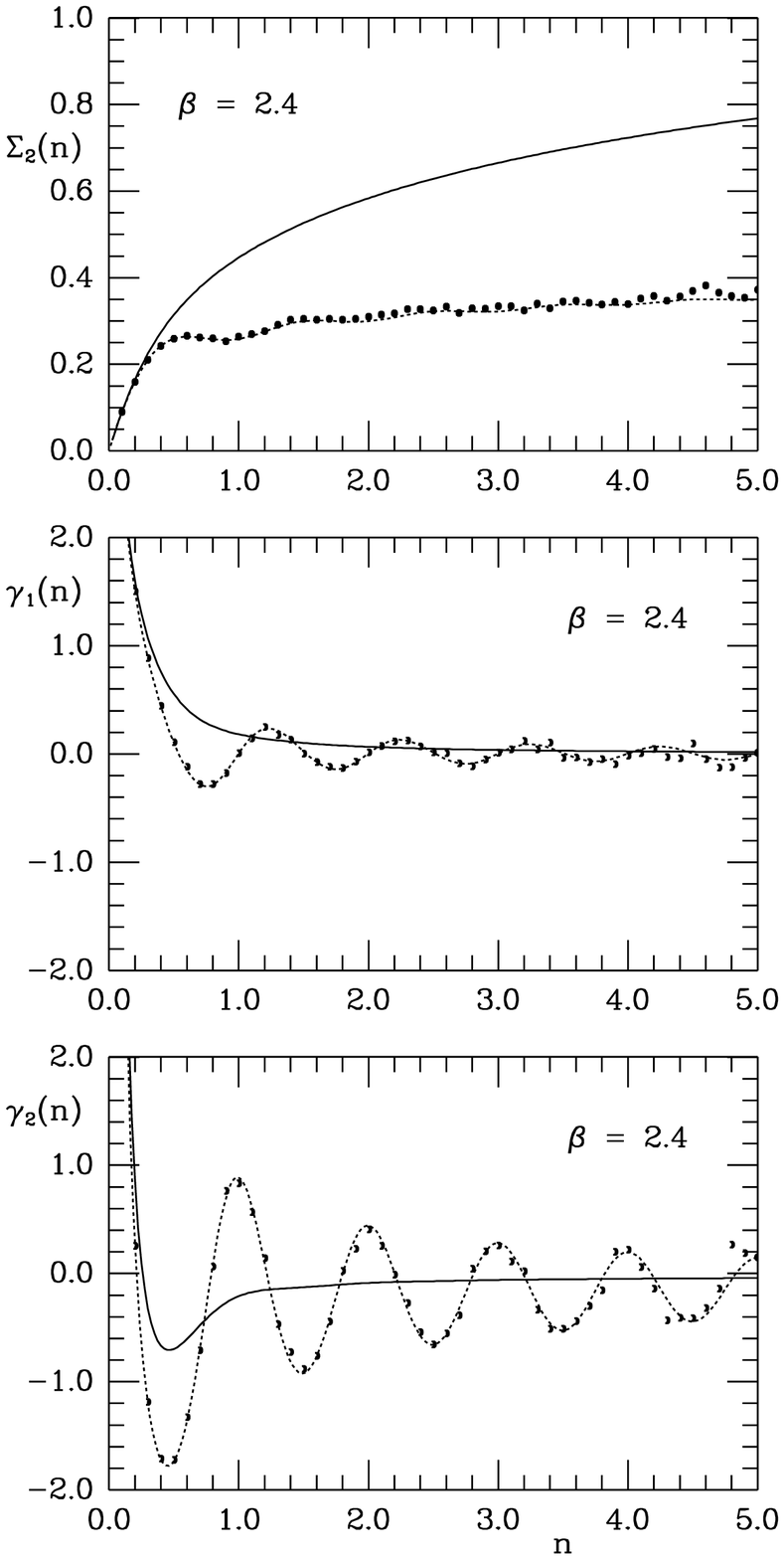,width=55mm}}}
\caption{The number variance, $\Sigma_2(n)$ and the first two cumulants,
$\gamma_1(n)$ and $\gamma_2(n)$ as a function of $n$.} 
\label{fig3}
\end{figure}
The first complete Dirac spectra on relatively large lattices were
obtained by Kalkreuter {\cite{Kalkreuter}} who
calculated $all$ eigenvalues of the $N_c = 2$ Dirac operator 
both for Kogut-Susskind (KS) fermions and Wilson fermions
for lattices as large as $12^4$.
In both cases, the Dirac matrix was tri-diagonalized by
Cullum's and Willoughby's Lanczos procedure {\cite{cullum}}
and diagonalized with a standard
QL algorithm.
This improved algorithm makes it possible to obtain $all$
eigenvalues. This allows us to test the accuracy of the eigenvalues
by means of sum-rules for the sum of the squares of the eigenvalues
of the lattice Dirac operator (see eqs. (\ref{sumks}, \ref{sumwil})). 
Typically, the numerical error in the
sum rule is of order $10^{-8}$.

Results for the spectral correlations for eigenvalues calculated by Kalkreuter
\cite{Kalkreuter}  both for KS and Wilson fermions are shown in
Fig. \ref{fig3}.
The simulations for KS fermions were performed  for 4 dynamical flavors
with $ma = 0.05$ on a $12^4$ lattice
(the lattice spacing is denoted by $a$). 
The simulations for Wilson fermions were
done for two dynamical flavors on an $8^3\times 12$ lattice.

The results for  $\Sigma_2(n)$, $\gamma_1(n)$ and $\gamma_2(n)$ obtained
by spectral averaging show an impressive agreement with the
RMT predictions. In the context of RMT it has been shown that 
spectral averages and ensemble averages coincide \cite{pandey}. This 
property is known as spectral ergodicity.

Spectra for different values of $\beta$ have been analyzed as well.
It is probably no
surprise that random matrix correlations are found at stronger couplings.
What is surprising, however, is that even in the
weak-coupling domain ($\beta =2.8$) the
eigenvalue correlations are in complete agreement with Random
Matrix Theory.

In the case of three or more colors with fundamental fermions, both the
Wilson and Kogut-Susskind Dirac operator do not possess any anti-unitary
symmetries. Therefore, our conjecture is that in this case   
the spectral correlations in the bulk of the spectrum of
both types of fermions can be described by
the GUE. This was recently confirmed for a wide range of
$\beta$-values both below and above the deconfinement phase
transition \cite{markum,Berg}.

As far as correlations in the bulk of the spectrum are 
concerned, in the case of two fundamental colors the continuum theory
and Wilson fermions are in the same universality class.
It is an interesting question of how spectral correlations of KS fermions
evolve in the approach to the continuum limit. Certainly, the
Kramers degeneracy of the eigenvalues remains. However, since Kogut-Susskind
fermions represent 4 degenerate flavors in the continuum limit,
the Dirac eigenvalues should obtain an additional two-fold degeneracy.
We are looking forward to more work in this direction.

\begin{figure}[!ht]
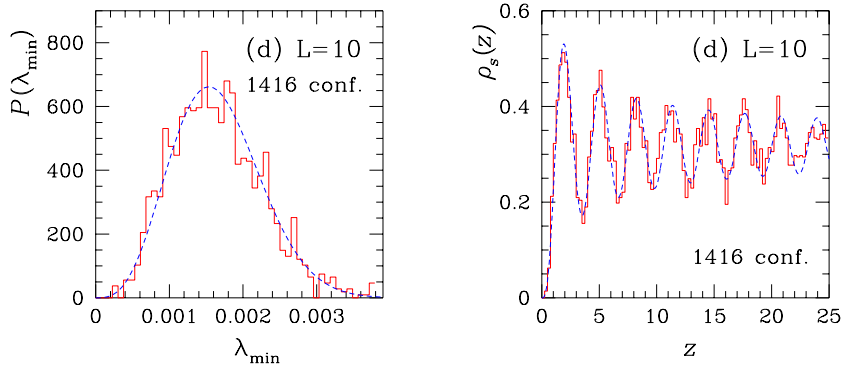

\vspace*{-1cm}
\hbox{
\psfig{file=tilo10small.ps,width=52mm,angle=0}
\hspace*{0.5cm}
\psfig{file=tilo10micro.ps,width=52mm,angle=0}}
\vspace*{-1.50cm}
\caption{The distribution of the smallest eigenvalue (left) and the
microscopic spectral density (right)
for two colors and $\beta = 2.0$.}
\label{fig4}
\end{figure}

  \subsection{Correlations near Zero Virtuality}

Spectral ergodicity cannot
be exploited in the study of the microscopic spectral density
and,  in order to gather sufficient statistics,
a large number of independent spectra is required.
One way to proceed is to generate instanton-liquid configurations
which can be obtained much more cheaply than lattice QCD configurations.
Results of such analysis {\cite{Vinst}} show that for $N_c=2$
with fundamental fermions the microscopic spectral density is given
by the chGOE. For $N_c =3$ it is given by the chGUE. One could argue that
instanton-liquid configurations can be viewed as
smoothened lattice QCD configurations. Roughening such configurations
will only improve the agreement with Random Matrix Theory.
Recently, these expectations were confirmed by lattice QCD simulations
\cite{Tilo}. Results for 1416
 quenched $SU(2)$ Kogut-Susskind
Dirac spectra  on a 10$^4$ lattice are shown in Fig. \ref{fig4}.
We show both the distribution of
the smallest eigenvalue (left) and the microscopic spectral density (right).
The results \cite{Nagao-Forrester}
for the chGSE are represented by the dashed curves.

Agreement of the microscopic spectral density with chRMT for $N_c=3$ was first
demonstrated by means of the valence quark mass dependence of
the chiral condensate \cite{vPLB}. Recently, these results were confirmed
by the calculation of complete Dirac spectra \cite{DamSU3,TiloSU3,Tilo99}. 
Other recent interesting results are the calculation of
the microscopic spectral density
for the Dirac operator in the adjoint representation \cite{Heller} 
and results for the distribution of the smallest Dirac eigenvalue at
nonzero topological charge obtained by means of the overlap 
formalism \cite{Kiskis}.

\begin{center}
\begin{figure}[!ht]
\centering\includegraphics[width=65mm,angle=270]{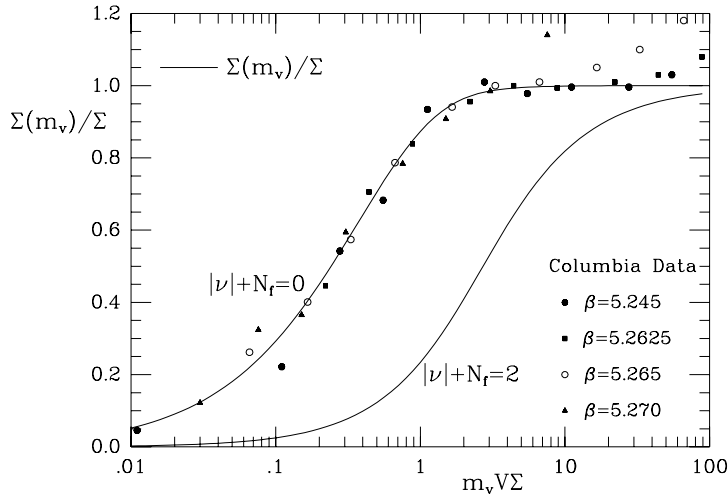}
\caption[]{The valence quark mass dependence of the chiral condensate
$\Sigma(m)$ plotted as $\Sigma(m_v)/\Sigma$ versus $m_v V\Sigma$. 
The dots and
squares represent lattice results by the Columbia group \cite{Christ}
 for values of $\beta$ as indicated in the label of the figure.}
\label{fig5}
\end{figure}
\end{center}

\subsection{The Valence Quark Mass Dependence of the Chiral Condensate}

The microscopic spectral density near zero virtuality was first studied
in terms of the valence quark mass dependence of the chiral condensate.
The lattice data for $\Sigma(m_v)$ were obtained by the Columbia
group \cite{Christ}
for  two dynamical flavors with  sea-quark mass $ma = 0.01$ and $N_c= 3$ on a
$16^3 \times 4$ lattice. 
In Fig. 4 we plot the ratio $\Sigma(m_v)/\Sigma$ as a function of 
$x=m_v V \Sigma$ (the 'volume' $V$ is equal
to the total number of Dirac eigenvalues), and compare the results
with the universal curves obtained from pq-QCD and chRMT (see eq. (\ref{val})).
 We observe
that the lattice data for different values of $\beta$ fall on a single curve.
Moreover, in the mesoscopic range
this curve coincides with the random matrix prediction for $N_f = \nu = 0$.
Of course this is no surprise.
In the region where valence quark mass is much less than
the current quark mass, the fermion determinant has no bearing on the Dirac
spectrum and we have effectively $N_f =0$.
On a lattice the zero mode states and a much larger
number of nonzero modes states are completely mixed, and therefore it is
not surprising that the effective topological charge that
reproduces the lattice results is equal to zero.

\section{Chiral Random Matrix Theory as a Schematic Model for the Dirac 
        Spectrum}
     \subsection{Introduction}
 Up to now we have mainly focussed on chiral Random Matrix Theory as an exact
theory for correlations of QCD Dirac spectra on the microscopic scale.
However, in the literature one finds another important application of 
RMT, namely as a schematic model for a disordered system. Two well-known
examples in this category are the Anderson model for localization 
\cite{Anderson} and
RMT applied to the theory of random surfaces in connection with quantum
gravity \cite{Ginsparg}. 
In this section we will study chiral random matrix theory
as a $schematic$ model of the chiral phase transition at nonzero 
temperature and chemical potential. 

      \subsection{chRMT at Nonzero Temperature}
In order to obtain a chiral random matrix model for QCD at nonzero 
temperature we first separate
 the Dirac operator in two pieces
\be
D = \tilde D + \gamma_0 \partial_0.
\ee
In a chiral basis with basis functions given by
\be
\psi_{kn}(x) = \phi_k(\vec x) e^{-\frac {2\pi i}\beta (n+ \frac 12) x_0}
\ee
the matrix corresponding to  $\gamma_0 \partial_0$ is block diagonal
with nonzero matrix elements given by the Matsubara frequencies. 
In our model we only keep  the lowest Matsubara frequencies $\pm \pi T$.
We expect that this is a reasonable approximation in the neighborhood
of the critical temperature and beyond. The matrix corresponding to
$\tilde D$ is replaced by a chiral random matrix.

The chRMT Dirac operator at $T\ne 0$ is thus given by \cite{JV,Tilo1}
\be
D^{RMT} = \left ( \begin{array}{cc} 0& iW \\ iW^\dagger &o\end{array} \right )
+ \left ( \begin{array}{cc} 0& i\Omega_T \\ i\Omega_T & 0 
\end{array} \right ),
 \ee
where $\Omega_T $ is given by 
\be
\Omega_T =\left ( \begin{array}{cc} \pi T {\bf 1}& 0\\
                                      0 & - \pi T {\bf 1}
\end{array} \right ),
\label{omegaT}
\ee
and ${\bf 1}$ is a unit matrix.

If we note that the probability distribution of the random matrix ensemble
is invariant under $W \rightarrow U W V^{-1}$ and use the equivalent of
$i\sigma_1 \sigma_3 \sigma_2 = 1$, the random matrix Dirac operator
can be simplified to
\be
 D(T)
 =\left ( \begin{array}{cc} 0 &i W+ \pi i T {\bf 1}\\
                                   i W^\dagger + \pi i T {\bf 1} &0
\end{array} \right ).
\label{DRMTT}
\ee
Notice that the dimension of the identity matrix in eqs.
(\ref{omegaT}) and (\ref{DRMTT}) differs by a factor 2.

Our finite temperature chiral random matrix model is thus given by
\be
Z(\nu,N_f) = \int dW {\det}^{N_f}( D(T) + m) 
e^{- n\Sigma^2 {\rm Tr} W W^\dagger}.
\ee 
In the next we will perform a saddle point analysis of this model.

         \subsection{Analytical Solution}

In this section we evaluate the partition function (\ref{ranpart}) using
methods which are standard in the supersymmetric formulation of random
matrix theory \cite{Efetov,VWZ}.  For simplicity we consider the model
for $\beta =2$ and choose a diagonal mass matrix with equal quark masses.
The first step is to perform
the average over $W$ by performing a gaussian integral. This leads
to a four-fermion interaction.  After averaging over the matrix
elements of the Dirac operator, the partition function becomes
\be
Z(\nu, N_f) = \int {\cal D} \psi^* {\cal D}\psi \exp[ &-&\frac
1{n\Sigma^2} \psi^{f\,*}_{L\, k}\psi^{f}_{R\, i} \psi^{g\,*}_{R\,
i}\psi^{g}_{L\,k }+{ m}\psi^{f\,*}_{R\, i} \psi^{f}_{R\, i}
+{ m}\psi^{f\,*}_{L\,
k}\psi^{f}_{L\, k}\nonumber \\
 &-&i\pi T(\psi^{f\,*}_{R\, i} \psi^{f}_{L\, i}
+\psi^{f\,*}_{L\,k}\psi^{f}_{R\, k})].
\ee
The four-fermion terms can be
written as the difference of two squares. Each square can be linearized by
\be
\exp(-AQ^2) \sim \int d\sigma\exp(-\frac{\sigma^2}{4A} - iQ \sigma) \ \ .
\label{Hubbard}
\ee
The different $Q$ 
variables, can be combined into a single complex $N_f\times N_f$ matrix, $A$,
resulting in
\be
Z(\nu, N_f) &=&
\int {\cal D} A
{\cal D} \psi {\cal D} \psi^* \exp[ -\frac{n\Sigma^2\beta}2 {\rm Tr}
A A^\dagger \nonumber\\ &-&i\psi^{f\,*}_{L\,k} 
\psi^{g}_{L\,k}(A +{ m}) -i\psi^{f\,*}_{R\,i} \psi^{g}_{R\,i}(A^\dagger
+{ m})
-i\pi T(\psi^{f\,*}_{R\, i} \psi^{f}_{L\, i}
+\psi^{f\,*}_{L\,k}\psi^{f}_{R\, k})] \ \ . \nonumber\\
\ee
Note that the temperature-dependent term can be rewritten as
\be
-\frac{i\pi T}2\left (\begin{array}{c}\psi_L\\ \psi_L^* \end{array}\right)
\left ( \begin{array}{cc} 0 & -{\bf 1}\\ {\bf 1} & 0 \end{array}
\right )
\left (\begin{array}{c} \psi_R \\ \psi_R^* \end{array}\right)
+ (L \longleftrightarrow R).
\ee
Using this, the fermionic integrals can be performed, and the
partition function with $|\nu| $ more left-handed modes than right-handed 
modes is given by
\be
Z(\nu, N_f) = \int {\cal D} A \exp [-\frac{n\Sigma^2\beta}2 {\rm Tr}
A A^\dagger] {\det}^{|\nu|} (A+{ m} ){\det}^{n}
\left ( \begin{array}{cc}  A+{ m} & -\pi i T \\  -\pi i T &A^\dagger +{
m}
\end{array} \right ).\nonumber \\
\label{apart2}
\ee
Here, $A$ is an arbitrary complex matrix and $m$ is proportional to the 
identity matrix.

The chiral condensate is defined by
\be 
\langle \bar \psi \psi \rangle = -\frac 1{2n N_f} \partial_{m} \log Z \ \ ,
\ee
For $n \rightarrow \infty$ it can be evaluated with the aid
of a saddle point approximation. The saddle point equations are given by
\be
-\frac{n\beta \Sigma^2}2 A+ n(A+m)\left( (A^\dagger +m)(A+m) + \pi^2 T^2\right
)^{-1} = 0 \ \ \ .
\label{spe}
\ee
An arbitrary complex matrix can be diagonalized by performing the
decomposition
\be
A= U \Lambda V^{-1},
\ee
with all eigenvalues positive and $U$ and $V$ unitary matrices.
We find that the solution of (\ref{spe}) yields $U = V = 1$ with
eigenvalues $\Lambda_k$ given by the positive roots of
\be
\Sigma^2\Lambda_k ((\Lambda_k+m)^2 + \pi^2 T^2) - \Lambda_k - m = 0  \ \ . 
\ee
In order to calculate the condensate, we express the derivative of the
partition function in (3.10) in terms of an average over $A$,
\be
|\langle \bar \psi \psi \rangle| =  
\frac 1{2n N_f} \langle{\rm Tr} \left (
\begin{array}{cc} A & -\pi iT \\ -\pi i T & A^\dagger \end{array}\right )^{-1}
\rangle \ \ .
\ee
Below $T_c$ and for $m\rightarrow 0 $ we find from the saddle point   
equation, 
\be
|\langle \bar \psi \psi \rangle| = \Sigma (1-\pi^2T^2\Sigma^2)^{1/2} \ \ .   
\label{cond1}
\ee
In the chiral limit we thus find a critical point at
\be
T_c = \frac 1{\pi \Sigma} \ \ \ .
\ee
Note that the dimensions are matched after including the mode density
$N/V =1$. Here, and elsewhere, our convention is that $N=2n$.
At $T_c$ the solution of the saddle point equation develops a non-analytic
dependence on $m$ resulting in the condensate
\be
|\langle \bar \psi \psi \rangle| = \Sigma^{\frac{4}{3}} m^{\frac{1}{3}} \ \ .
\ee
Therefore, we reproduce the mean field value for the critical exponent
$\delta = 3$.

It is also possible to obtain an analytical expression for the spectral
density of this model as a function of the critical temperature. It can
be obtained by solving the Schwinger-Dyson equations 
for the resolvent of the model \cite{JV} or 
from the observation that the solution of (\ref{spe}) does not depend on the 
number of flavors. For $N_f \rightarrow 0$ the spectral density is then
obtained from the discontinuity of the chiral condensate across the
real $m$-axis \cite{Stephanov1}. 
The spectral density thus follows from the solution of a cubic
equation. The zero temperature limit is a semicircle whereas the high
temperature limit is given by two semi-circles separated by $2\pi T$.

      \subsection{chRMT at Nonzero Chemical Potential}

The chemical potential enters in the continuum QCD partition function as
\be
i\gamma_0 A_0 \rightarrow i\gamma_0 A_0 + \mu \gamma_0.
\ee
 resulting in the chRMT Dirac operator \cite{Stephanov}
\be
D(\mu) = \left ( \begin{array}{cc} 0 & iW + \mu \\ iW^\dagger + \mu & 0
\end{array} \right ).
\label{Dmu}
\ee
If we notice that $\mu \gamma_0$ has the same reality properties as 
$\gamma_0 \partial_0$, we immediately conclude \cite{Osbornmu}
that these ensembles
can be classified according to the same anti-unitary symmetries as
the chRMT's for $\mu = 0$.
However, the term proportional to $\mu$ violates the anti-Hermiticity of the 
Dirac operator, and typically the eigenvalues of $D(\mu)$
will be scattered in the
complex plane.

Many of the problems associated with the presence of a chemical potential
are related to the loss of non-Hermiticity. Exactly, these problems
are reproduced by the chRMT partition function with Dirac operator given
by (\ref{Dmu}). For example, in this model 
one can study the problems of the quenched
approximation \cite{everybody,Klepfish,barbour}, 
the structure of the Yang-Lee zeros \cite{Barbourqed,Halaszyl}
and the problems with the Glasgow method \cite{Halaszgl}. 
Below we will discuss the first two applications. 

          \subsection{Yang-Lee Zeros}

The QCD partition function 
\be Z(m,\mu)= \int {\det}^{N_f} (D + m + \mu\gamma_0) e^{-S_{YM}}
\ee
is a polynomial in $m$ and $\mu$. For simplicity, let us consider the
case $m= 0$. Then \cite{Vink}
\be
Z^{QCD}(m=0,\mu) \sim \prod_k(\mu -\mu_k),
\ee
and the baryon density is given by
\be
n_B = \frac 1V\partial_\mu \log Z(m=0,\mu) = \frac 1V\sum_k \frac 1{\mu-\mu_k}.
\ee
With the zeros scattered in the complex $\mu$-plane we can 
interpret the real and imaginary parts of $n_B$ as the electric
field at $\mu$ due to charges located at $\mu_k$. 

Physically, we expect that $n_B= 0$ for $\mu <\mu_c \ne 0$ and
jumps to a finite value at $\mu= \mu_c$. We thus expect a first 
order phase transition at $\mu_c$. Two possibilities for the
distribution of the zeros come to mind. First,
according to Gauss law, a homogenous
distribution of zeros $\mu_k$ along the complex circle $|\mu| = \mu_c$
results in a zero baryon number density for $|\mu  | < \mu_c$.
(Of course in the thermodynamic limit, a subleading number of zeros
may be present inside the circle.). Second, we all know that
the electric field is zero between
two infinite parallel plates with equal constant charge density is zero.
The analogue of this phenomenon in two dimensions is that the electric 
field is zero between
two parallel infinite line charges with constant charge density.
Therefore, the second possibility is that the zeros in the complex $\mu $-plane
are located at
\be
\mu_k = \pm \mu_c + \frac {k\pi i} {2\beta}, \qquad k \,\,{\rm odd}.
\label{zeroline}
\ee
We also expect that the baryon 
number density increases smoothly for  $\mu >\mu_c$. This requires 
a radially symmetric cloud of zeros outside of the radius $|\mu| = \mu_c$ in
the first case, and a constant density parallel to the imaginary axis for
$|{\rm Re} (\mu)|>\mu_c$ in the second case.

The simplest possible model for spherically symmetric distributed
zeros of the partition
function, is one with zeros given by the roots of unity,
\be
\mu_k = \mu_ce^{\frac{2\pi i k}{N}}, \quad \,\, k = 1, \cdots, N  .
\ee
This results in the partition function
\be
Z(\mu) = \prod_k (\mu -\mu_k) = \mu^N - \mu_c^N.
\ee
For the baryon number density we then obtain
\be
n_B &=& \frac 1N \partial_\mu \log Z = \frac {\mu^{N-1}}
{\mu^N - \mu_c^N}\nonumber \\
    &=& \left \{ \begin{array}{c} \frac 1\mu \quad {\rm for} \quad|\mu | > 
\mu_c \\
                 0 \quad{ \rm for} \quad |\mu | < \mu_c \end{array} \right .
\quad {\rm for} N\rightarrow \infty .
\ee

In the second case with zeros given by (\ref{zeroline}), the partition function
is given by
\be
Z(\mu) = \prod_{k\,\, {\rm odd}} (\mu - \mu_c - \frac{k\pi i}{2\beta})
(\mu +\mu_c- \frac{k\pi i}{2\beta}).
\ee
Up to a constant the zeros are given by the zeros of the $\cosh$-function.
We thus find that
\be
\frac {Z(\mu)}{Z(0)} = \frac{\cosh(\beta(\mu_c -\mu)) \cosh(\beta(\mu_c+\mu))}
{\cosh^2(\mu_c)}.
\ee
The baryon density is then given by (in an arbitrary normalization)
\be
n_B = \frac 1{2\beta} \partial_\mu \log Z(\mu) = \frac {\sinh(2\beta \mu)}
{\cosh(2\beta \mu) + \cosh(2\beta \mu_c)}.
\ee
For $\beta \rightarrow \infty$ this can be approximated by
\be
n_B = \frac 1{1+e^{2\beta(\mu_c -\mu)}},
\ee
and we find that 
\be
 n_B = 0 \quad {\rm for} \quad \mu < \mu_c,\\
 n_B = 1 \quad {\rm for} \quad \mu > \mu_c.
\ee
In the next section we will study the baryon number density in the 
chiral random matrix model at nonzero chemical potential. 

\subsection{Phases in chRMT at $\mu \ne 0$}

The chRMT partition function at $\mu \ne 0$ is obtained from the finite
$T$ partition function by the replacement 
$T \rightarrow i\mu$. For  $N_f = 1$ this results in the $\sigma$ model 
\be
Z(\mu) =\int d\sigma d\sigma^* (|\sigma|^2 - \mu^2)^N e^{-N|\sigma|^2}.
\ee
For $N\rightarrow \infty$, the integrals in this partition partition
function can be performed by a saddle point approximation. We find that
\be
\bar \sigma = 0 \quad \hbox{for} \quad \mu >\mu_c \,\,\,
&\Rightarrow Z& = \mu^{2N}, \\
\bar \sigma = \sqrt{1+\mu^2}\quad  \hbox {for} \quad \mu < \mu_c \,\,\,
&\Rightarrow& Z = e^{-N(\mu^2 + 1)}.
\ee
The critical point is given by the trancendental equation \cite{Stephanov}
$\mu_c^2 = 
\exp(-1-\mu_c^2)$ which is solved by  $\mu_c \approx 0.53$
  
\begin{center}
\begin{figure}[!ht]
\centering\includegraphics[width=45mm]{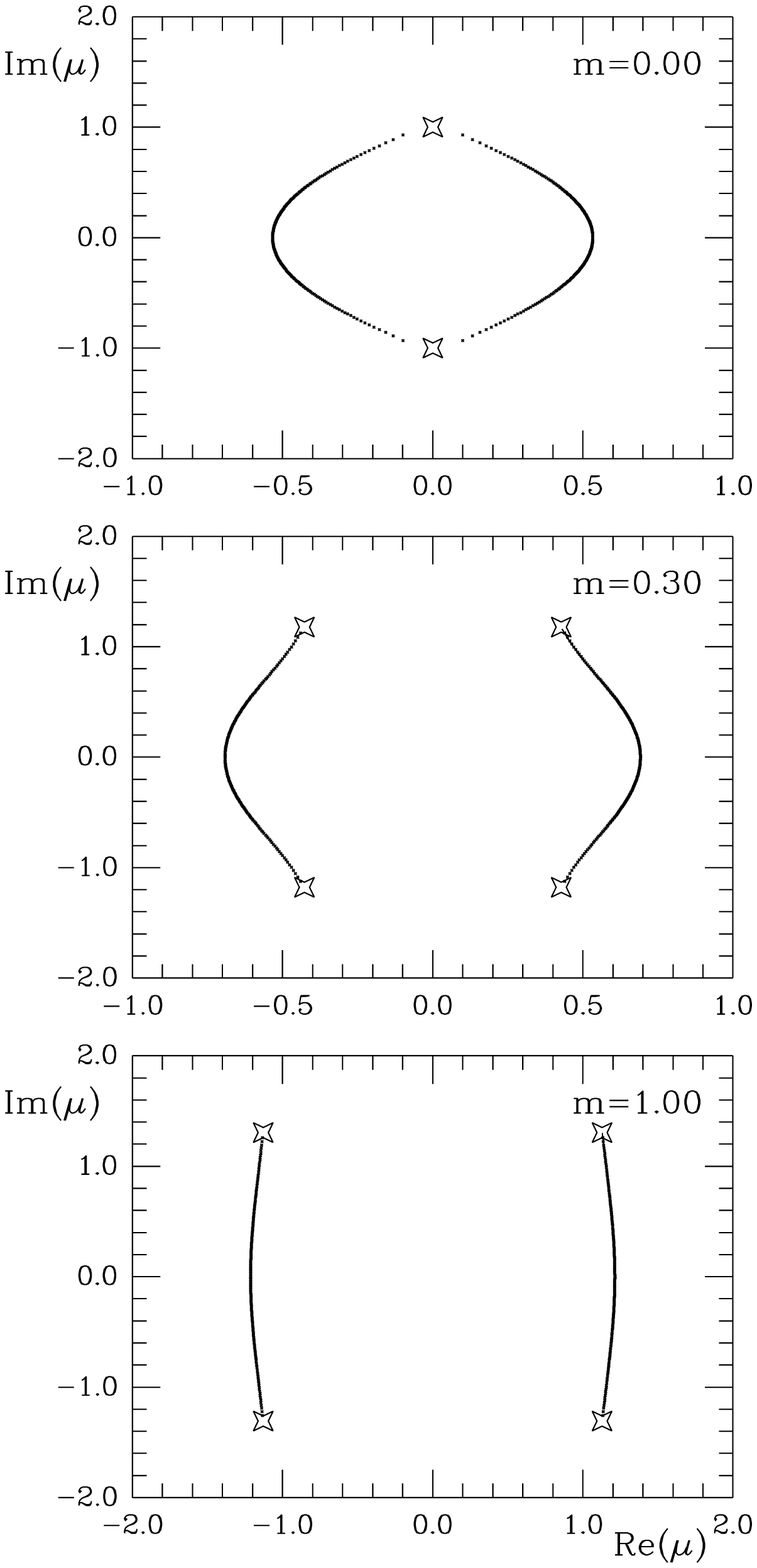}
\includegraphics[width=45mm]{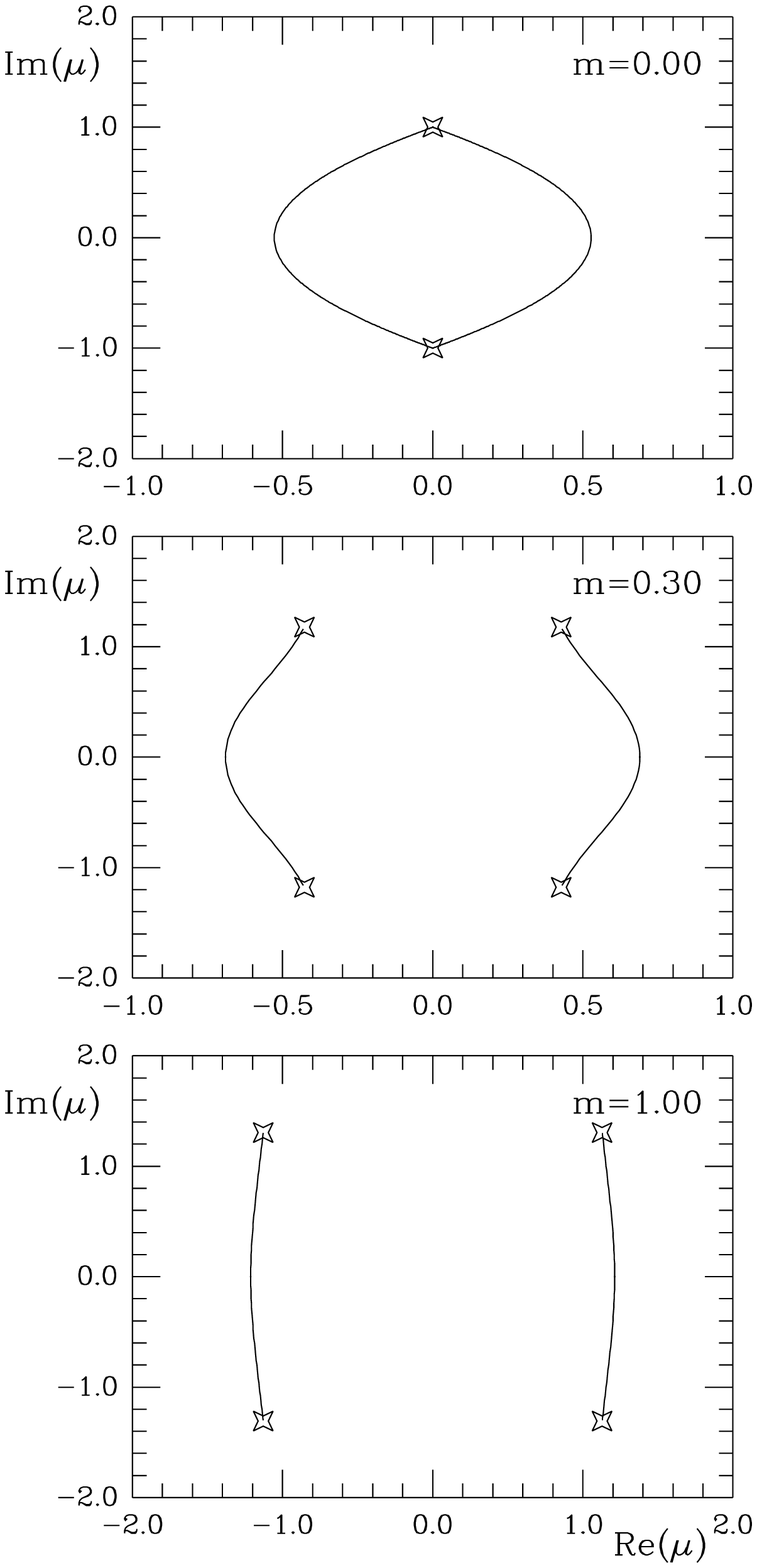}
\caption[]{
 The zeros of the partition function
in the complex $\mu$ plane (left) and the result obtained from a
mean field analysis (right).
Results are given for
 for $m =0$ (upper), $m = 0.30$ (middle), and $m = 1.0$ (lower) for
$N = 192$. 
The zeros of the discriminant
of the cubic equation  are denoted by stars.  Note that the
scale on the $x$-axis of the lower figure is different.}
\end{figure}
\end{center}
We find that the baryon number density above the critical point 
shows the same behavior as in our naive model. However,
the baryon density for $\mu < \mu_c$ is nonvanishing which disagrees
with our expectation for QCD at zero temperature and finite density.
The remedy should be clear. In order to obtain a sharp
Fermi-Dirac distribution one has to sum over all
Matsubara frequencies \cite{Kapusta}. This 
point was ignored in our model. In the next section we discuss a RMT 
$\sigma$-model that explicitly includes the lowest two Matsubara frequencies.
This model can be trivially extended to include all Matsubara frequencies, 
and the infinite product can be evaluated analytically in terms of a
$\cosh$-function. For $\beta \rightarrow \infty$ we find \cite{JVmu}
that the zeros
of the partition function in the complex $\mu$-plane are located on the
line ${\rm Re}(\mu) = \pm \mu_c$.

A much more disturbing observation is that in the thermodynamic limit the
partition function for $\mu < \mu_c$ becomes 
exponentially small \cite{Halaszyl,Halasz}. 
This exponential cancellation is due to the phase of the eigenvalues.
It makes accurate numerical simulations forbiddingly difficult.

The random matrix partition function is a polynomial in $m$ and $\mu$ and
it is straightforward to study the zeros of the partition function in 
the complex $m$-plane or the complex $\mu$-plane. To assure our numerical 
accuracy we have determined the zeros of the polynomials 
by means of  multi-precision
algorithms \cite{bailey}
using as much as 1000 significant digits. The results are
shown in Fig. 5. The branch points at the end of a cut are the points
where two saddle-point solutions coincide. In the present case they
are given by zeros of the discriminant of a cubic equation.
The line of zeros can also be obtained from a saddle point 
analysis (see right half of the figure). For more
details and results in other fields \cite{Frank,Shrock}
we refer to the original literature.

          \subsection{Failure of the Quenched Approximation}

For QCD with three or more colors and fermions in the fundamental 
representation, the fermion determinant is complex. The presence of
this phase makes Monte-Carlo simulations impossible. A way out might
be the quenched approximation, i.e. ignore the fermion determinant
altogether and hope that it will work. This approach was followed
in \cite{everybody,Klepfish} with the conclusion that the critical chemical
potential is given by the pion mass instead of the nucleon mass. Obviously, 
this result is physically wrong!

This problem was first analyzed in chRMT by Stephanov \cite{Stephanov}. 
He showed
that the quenched limit is the limit $N_f \rightarrow 0$ of a partition
functions with fermion determinant
\be
|\det(D(\mu) +m)|^{N_f}
\ee
rather than
\be
(\det(D(\mu) +m))^{N_f}.
\ee

The absolute value can be represented as (let us take $N_f =2$ for simplicity)
\be
\det(D(\mu) +m)
\det(D^\dagger(\mu) +m^*)= \int D\psi D\psi^c e^{\bar\psi(D(\mu) + m) \psi +
\bar\psi^c(D^\dagger(\mu) + m^*) \psi^c},\nonumber \\
\label{conjugate}
\ee
 and Goldstone bosons made out of quarks and conjugate anti-quarks
carry a net baryon number resulting in a critical chemical
potential given by the pion mass.

The random matrix model corresponding to (\ref{conjugate}) can be mapped
onto a nonlinear $\sigma$-model which can be solved by a saddle-point
approximation. The exact solution of this model confirms that 
the critical chemical potential is proportional to $\sqrt m$, the same
dependence as for the pion mass.

      \subsection{The Chiral Phase Transition in chRMT}
In previous section we have studied a model at nonzero temperature and
a model at nonzero chemical potential. These two models can be combined in 
the following schematic chRMT model \cite{phase} for the chiral phase 
transition,
\be
Z = \int DW {\det}^{N_f} \left ( \begin{array}{cc} m & iW + iC\\
                                 iW^\dagger +iC& im \end{array}\right )
\ee
with $C$ a diagonal matrix with $C_k=a\pi T
-ib\mu$ for one half of the diagonal elements  and
$C_k=-a\pi T-ib\mu$ for the other half with $a$ and $b$
dimensionless parameters.
This model can be considered as the matrix equivalent of a Landau-Ginzburg
functional. We thus expect to find mean field critical exponents. The advantage
over using Landau-Ginzburg theory is that in this case the spectrum of
the Dirac operator is accessible. In particular, this might reveal interesting
behavior of the Dirac spectrum near the critical point. For example, 
for $\mu = 0$ we have shown that the fluctuations of the smallest eigenvalue
of the Dirac operator can be used as an order parameter for the chiral
phase transition \cite{JV}. 

Also in this case the partition function can be reduced to a $\sigma$-model,
For the simplest case of $N_f = 1$ it is given by
\be
Z(T,\mu) = \int d\sigma e^{-N\Omega (\sigma)},
\ee
where
\be
\Omega(\sigma) = \sigma \sigma^\dagger
&-&\log ((\sigma+m)(\sigma^\dagger + m) - (\mu+iT)^2)\nonumber \\
&-&\log ((\sigma+m)(\sigma^\dagger + m) - (\mu-iT)^2).
\ee
The chiral condensate is given by the expectation value of $\sigma$. The 
saddle-point equation is a fifth order equation in $\sigma$. Therefore this
model is very similar to a $\phi^6$ Landau-Ginzburg theory. For  $m=0$ we 
find that $\sigma$ is real and the saddle point-equation is given by
\be
\sigma[\sigma^4 - 2(\mu^2 -T^2 +\frac 12)\sigma^2 + (\mu^2+T^2)^2 +\mu^2 -T^2]
 =0.
\ee
The critical points occur where one of the solutions of the cubic equation
merge with the solution $\sigma = 0$, i.e. along the curve
$(\mu^2+T^2)^2 +\mu^2 -T^2  =0$. At the tri-critical point three solution
merge. This happens if in addition $\mu^2 -T^2 +\frac 12=0$.

In Fig. 6 we show the phase diagram in the $\mu T m$ space. In the $m=0$
plane we observe a line of second order phase transitions and a
line of first order phase transitions. They join at the tricritical point. 
Also joining at the tricritical point is a line of second order
phase transitions in the $m$-directions which is the boundary of the 
plane of first order transitions in $\mu Tm$-space.

\begin{figure}[htb]
\setlength{\unitlength}{2.4in}
\centerline{\psfig{file=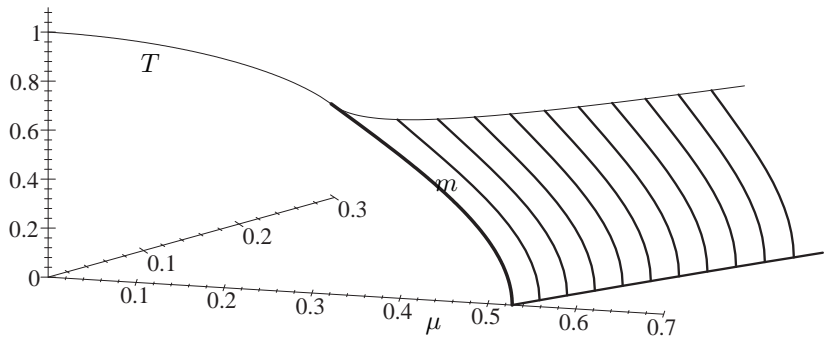,width=2\unitlength}}
\vspace{-\unitlength}
\begin{picture}(2,1)
\put(.34,.62){ $T$}
\put(.96,.06){ $\mu$}
\put(1.00,.36){$m$}
\end{picture}
\caption[]{\small Phase diagram of QCD with two light flavors
of mass $m$ as calculated from the random matrix model.
The almost parallel curves on the wing surface
are cross sections of this surface with $m=$const planes.
}
\label{fig:3dRM}
\end{figure}

            \subsection{Possibility of a Localization Transition in QCD}    
 From the theory of Anderson localization we know that the eigenvalues
corresponding to localized states are statistically independent and 
thus obey Poisson statistics. Since it has been well established that
in QCD at zero temperature  the correlations of the eigenvalues 
of the QCD Dirac operator are given by
chiral Random Matrix theory, we conclude that the eigenstates are extended.
This also follows from a direct analysis of the Dirac wave functions for
gauge field configurations given by a liquid of instantons. However, we wish
to point out that studies with Wilson Dirac fermions
indicate that eigenfunctions are localized \cite{Janssen}. We do not
have an explanation for this discrepancy.

One reason to expect a localization transition with increasing  temperature is
dimensional reduction. States are more likely to be localized in lower
dimensions. For example, in a one-dimensional disordered system all
states are localized. One the other hand, because of asymptotic freedom
the coupling becomes weaker at higher temperatures.
Let us analyze more closely what will happen.

Lattice simulations for the valence quark mass dependence of the chiral
condensate were performed for temperatures both below and above the
critical temperature \cite{Christ}. It was found that, in the ergodic
domain, all data can be rescaled onto a universal curve given by
chRMT. This shows that the states are extended below the critical point.

According to a theoretical argument due to Parisi \cite{Parisi}
given in the 
context of disordered systems a localization transition can only occur in
quenched theories. In QCD, this argument can be translated as follows.
Let us assume that the eigenfunctions of the Dirac operator are localized.
Then the joint eigenvalue distribution factorizes in one-particle distributions
$(\lambda^2 +m^2)^{N_f}F(\lambda)$.
The average spectral density is thus given by
\be
\rho(\lambda_1) &=& \int d\lambda_2 \cdots d\lambda_n \prod_{k=1}^n 
(\lambda_k^2 +m^2)^{N_f} F(\lambda_1) \cdots F(\lambda_n)  \nonumber \\
&=& c_1(\lambda_1^2 +m^2)^{N_f} F(\lambda_1).
\ee
We then find
\be
\lim_{\lambda_1 \rightarrow 0}\lim_{m\rightarrow 0}  
\lim_{V\rightarrow \infty} 
\rho(\lambda_1) = 0,
\ee
i.e., no spontaneous breaking of chiral symmetry.
We thus conclude that in the chiral limit no localization can occur
in QCD with light quarks. 
What happens in quenched theories remains an open question. However, 
studies with long range interactions given by $\sim \pm 1/|R_i - R_j|^d$
with random signs and random positions $R_i$ as is the case for instanton
liquid simulations indicate that all states are delocalized \cite{Parshin}.

          \subsection{Triality at $\mu \ne 0$}
Let us finally discuss how Dirac spectra at nonzero chemical potential
depend on the Dyson index of the matrix elements. We remind the reader
that the classification of Dirac operators with chemical potential is
the same as for zero chemical potential. 

Numerical simulations have been performed for all three classes.
A cut along the imaginary axis below a
cloud of eigenvalues was found in instanton liquid
simula\-tions {\cite{Thomas}}  for $N_c =2$ at $\mu \ne 0$
which corresponds to $\beta =1$. In lattice QCD simulations
with staggered fermions for $N_c = 2$ {\cite{baillie}}
a depletion of eigenvalues along the imaginary axis was observed, whereas
for $N_c=3$ the eigenvalue distribution did not show any pronounced
features \cite{everybody}.

\begin{figure}[!ht]
{\makebox{\epsfig{file=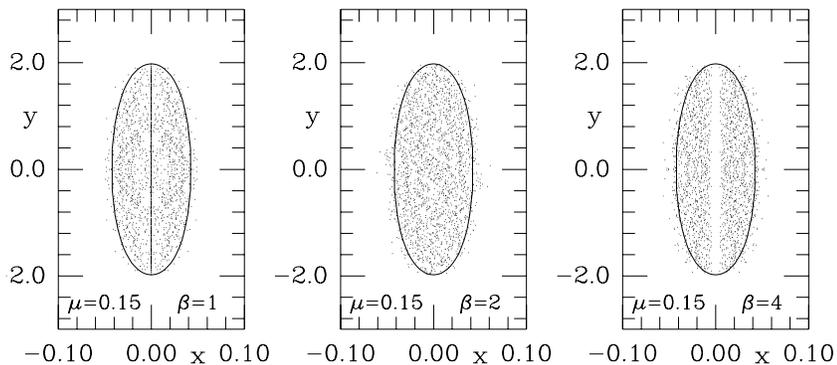,
width=110mm,angle=0}}}
\caption[]{
Scatter plot of the real ($x$), and the imaginary
parts ($y$) of the eigenvalues of the random matrix Dirac operator
at nonzero chemical potential.
The values of $\beta$ and $\mu$ are given in the labels of the figure.
The full curve shows the analytical result for the boundary.}
\label{fig6}

\end{figure}

In the quenched approximation, the spectral properties
of the random matrix Dirac operator (\ref{Dmu})
can easily be studied numerically by  diagonalizing a set of
matrices with probability distribution (\ref{ranpart}).
In Fig. 7  we show results {\cite{Osbornmu}}
for the eigenvalues of a few
$100\times  100$ matrices for $\mu = 0.15$ (dots). The solid curve represents
the analytical result for the boundary of the domain of eigenvalues
derived in {\cite{Stephanov}} for $\beta =2$. However,
the method that was used can be extended {\cite{Osbornmu}}
to $\beta = 1$ and $\beta =4$ and with the proper scale factors we find
exactly the same solution.

For  $\beta =1$ and $\beta = 4$ we observe exactly the same structure as in
the previously mentioned (quenched) QCD simulations.
We find an accumulation of eigenvalues on the imaginary axis for $\beta = 1$
and a depletion of eigenvalues along this axis for $\beta = 4$.
This depletion can be understood as follows. For $\mu = 0$ all eigenvalues
are doubly degenerate. This degeneracy is broken at $\mu\ne 0$ which produces
the observed repulsion between the eigenvalues.

The number of purely imaginary eigenvalues for $\beta = 1$
scales as $\sqrt N$ and is thus not visible in
a leading order saddle point analysis.
Such a $\sqrt N$  scaling is typical
for the regime of weak non-hermiticity first identified by Fyodorov
{\it et al.} {\cite{fyodorov}}.
Using the supersymmetric method of random matrix theory
the $\sqrt N$ dependence was obtained
analytically by Efetov {\cite{Efetovnh}}.
Also the case $\beta=4$ was analyzed
analytically in \cite{efetovsym} with results that are in complete agreement
with our numerical simulations.
Obviously, more work has to be done in order to
arrive at a complete characterization of
universal features {\cite{fyodorovpoly}} in
the spectrum of nonhermitean matrices.

\section{Closing Remarks}
In these lectures we have presented analytical results for 
the infrared limit of the QCD Dirac spectrum. We have shown
that the correlations of the Dirac eigenvalues for level spacings below
the Thouless energy are given by the zero momentum limit of the partially
quenched chiral partition function and agree with results obtained from
chiral random matrix theory. A key ingredient in the formulation of the
effective theory is the structure of the integration manifold. We have 
argued that it is given by a super-Riemannian manifold which is characterized
by a symbiosis between compact and noncompact degrees of freedom.
We have shown that the valence quark mass dependence of the chiral condensate
that follows from the zero momentum sector of the effective 
chiral partition function coincides with the result from chiral Random 
Matrix Theory. In this formulation, universality is natural. The structure
of the chiral Lagrangian only depends on the pattern of chiral symmetry
breaking.
 
An important condition for the validity of the chiral Lagrangian 
is the requirement that the only low-lying modes are the Goldstone modes
associated with the spontaneous breaking of chiral symmetry. Without
confinement this would not be the case. 
On the other hand, according to the Bohigas conjecture, eigenvalue correlations
are given by random matrix theory if the corresponding classical system
is chaotic. If the classical motion of quarks in 4+1 dimensions is 
chaotic the eigenvalues of the Dirac operator are correlated 
according to chRMT. Then necessarily the low-energy dynamics is given
is given by the chiral Lagrangian and this can only happen if we
have confinement. 

Can we reverse this statement? Is the classical motion of a fermion in
a confining theory necessarily chaotic? 
The point I wish to make is that the Bohigas conjecture and confinement 
are not unrelated. For a deeper understanding of confinement a proof of
the Bohigas conjecture might be required.  Recent progress in this
direction is encouraging \cite{andreev,Martinkick,ast,Martinap,martinbgs}.

\section{Acknowledgments} I wish to thank the organizers of the Osako
school and the Kyoto workshop for their hospitality and for organizing
such a wonderful meeting. In writing this review 
I have benefitted greatly from discussions and  collaborations with 
Gernot Akemann, Igor Aleiner, Alexander Altland, Poul Damgaard, 
Yan Fyodorov, Meinulf G\"ockeler
Thomas Guhr, Adam Halasz, Bertram Klein, Andy Jackson,
Steffen Meyer,
Shinsuke Nishigaki, James Osborn, Paul Rakow, Andreas Sch\"afer, 
Melih Sener, Ben Simons, Robert Shrock,
Edward Shuryak,
Andrei Smilga, Misha Stephanov, Dominique Toublan, Hans Weidenm\"uller,
Tilo Wettig and 
 Martin Zirnbauer. 
Dominique Toublan is acknowledged for a critical reading of the
manuscript.

\end{document}